\documentclass[12pt]{article}
\usepackage{amssymb,amsmath,epsfig,mathtools,caption}

\begin{document}

\title{\bf Thermodynamics of Black holes With Higher Order Corrected Entropy}
\author{M. Umair Shahzad$^{1}$\thanks{m.u.shahzad@ucp.edu.pk}
and Abdul Jawad$^2$\thanks{jawadab181@yahoo.com; abduljawad@cuilahore.edu.pk} \\
\small $^1$Center for Applicable Mathematics and Statistics, Business School, \\
\small  University of Central Punjab, Lahore, Pakistan.\\
\small $^2$Department of Mathematics, COMSATS University Islamabad,\\
\small Lahore Campus, Pakistan}

\date{}
\maketitle

\begin{abstract}

For analyzing the thermodynamical behavior of two well-known black
holes such as RN-AdS black hole with global monopole and $f(R)$
black hole, we consider the higher order logarithmic corrected
entropy. We develop various thermodynamical properties such as,
entropy, specific heats, pressure, Gibb's and Helmhotz free energies
for both black holes in the presence of corrected entropy. The
versatile study on the stability of black holes is being made by
using various frameworks such as the ratio of heat capacities
($\gamma$), grand canonical and canonical ensembles, and phase
transition in view of higher order logarithmic corrected entropy. It
is observed that both black holes exhibit more stability (locally as
well as globally) for growing values of cosmological constant and
higher order correction terms.

\end{abstract}

\section{Introduction}

A classical black hole (BH) resembles an object in the thermodynamic
equilibrium state which was firstly observed by Bekenstein \cite{1},
that led a concept of BH entropy. In this way, Hawking \cite{2}
discovered heat emission from BHs and introduced the famous formula
of entropy, i.e., it is proportional to area of event horizon. The
event horizon only allows us to explore the information about mass,
charge and angular momentum (no hair theorem) \cite{3}. So, various
BHs have same charge, mass and angular momentum which is formed by
different configuration of in-falling matter. These variables are
similar to the thermodynamical variables of pressure and energy.
There are many possible configurations of a system leading to the
same overall behavior. This leads to the concept of maximum entropy
of BHs \cite{4} which needs corrections because of quantum
fluctuations and paved the way of holographic principle \cite{7,8}.
These fluctuations lead to the corrections of standard relation
between area and entropy because BH size is reduced due to Hawking
radiations \cite{9}.

In statistical mechanics, there are the notions of the canonical and
the microcanonical ensembles. In general, we can define entropy in
canonical as well as microcanonical ensembles. There is the
difference between two definitions which should be noted. The energy
is allowed to fluctuate about a mean energy $\bar{E}$ in the
canonical ensemble, while the energy is fixed (say at point $E$) in
the microcanonical ensemble. If one uses the entropy as information
then the canonical entropy must be higher than the microcanonical
entropy. This is because there is an additional ambiguity that which
configurations a system can take as the system can be in a
configuration with energy close to $\bar{E}$ apart from the
configurations with energy equal to $\bar{E}$ in the canonical
ensemble \cite{4}. The leading order difference between the
microcanonical and canonical entropies for any thermodynamic system
is given by $- 1/2 ln(C T^2)$ where $C$ and $T$ is the specific heat
and temperature of the system \cite{9a}. This is accounted for by
the logarithmic corrections found by other methods \cite{9b}
suggesting that the semiclassical arguments leading to Bekenstein
Hawking entropy give us the canonical entropy. That such a
correction to the canonical entropy leads closer to the
microcanonical entropy also has been verified analytically and
numerically in \cite{9a}.

Higher order corrections to thermodynamic entropy occur in all
thermodynamic systems when small stable fluctuations around
equilibrium are taken into account. These thermal fluctuations
determine the prefactor in the density of states of the system whose
logarithm gives the corrected entropy of the system. Higher order
corrections to Bekenstein standard area entropy relation can be
interpreted as corrections due to small fluctuations of BH around
its equilibrium configuration which is applicable to all BHs with
positive heat capacity \cite{9a}. This analysis simply uses
macroscopic properties (entropy, pressure, etc.) but does not use
properties of microscopic theories of gravity such as quantum
geometry, string theory, etc. It is argued that results of \cite{4}
are applicable to all classical black holes at thermodynamics
equilibrium. However, one can use the first and second order of
corrections in some unstable black holes under the special
conditions. In that case the approximation is valid as one consider
only small fluctuations near equilibrium. But we should comment that
as the black hole becomes really small possible near Planck scale
such approximation should break down which can not be trusted. But
as long as we consider small fluctuations we can analyze them
perturbatively, and therefore we consider only the first and second
order corrections and neglect the higher order terms in the
perturbative expansion \cite {16a}.

There are various approaches to evaluate such corrections
\cite{10,14}. For example, thermal fluctuations effects have been
analyzed onto different BHs by using logarithmic correction upto
first order \cite{9,15,16}. However, it is also possible to evaluate
higher order correction to BH entropy for analyzing the thermal
fluctuations around the equilibrium \cite{4}. It is argued that the
results of higher order corrections are applicable to all classical
BHs at thermodynamics equilibrium. Moreover, these corrections has
been applied to different BHs \cite{16a}. However, one can utilize
the first (logarithmic) as well as second order correction on some
unstable BHs. These corrections are valid and can apply on small
fluctuations near equilibrium \cite{16a}.

There is a strong observational evidence such as type 1a supernova
\cite{16b}, cosmic microwave background \cite{16c} and large scale
structure \cite{16d,16e} have shown that 'dark energy' dominates the
energy budget of the Universe. The dynamical effect of dark energy
is responsible for the accelerated expansion of Universe. Recently,
the simplest best strategy is to model the dark energy via small but
positive cosmological constant \cite{16f}. Furthermore, during the
phase transitions, many topological defects are produced such as
domain walls, cosmic strings, monopoles, etc. Monopoles are the
three-dimensional topological defects that are formed when the
spherical symmetry is broken during the phase transition \cite{17}.
The pioneer work in this direction was done by Barriola and Vilenkin
\cite{18}, who found the approximate solution of the Einstein
equations for the static spherically symmetric BH with a global
monopole (GM). Many authors have also investigated different
physical phenomena of BHs with GMs \cite{19,20,20a}.
 However, thermodynamical properties of BHs with GMs still remain
obscure, although it deserves a detail analysis of thermodynamics.
One of the modified theories is $f(R)$ theory of gravity
\cite{21,22,23,24,24a}, which explains the accelerated expansion of
the universe. There are a lot of valid reasons because of which
$f(R)$ gravity has become one of the most interesting theories of
this era. It is important and interesting to study the astrophysical
phenomenon in $f(R)$ gravity \cite{25}.

In this work, we will utilize the higher order corrected entropy for
evaluating various thermodynamical properties for RN-AdS with GM and
$f(R)$ BHs. Rest of the paper can be organized as: In section
\textbf{2}, we discuss the thermal fluctuations by utilizing the
higher order correction terms. In section \textbf{3} and \textbf{4},
we analyze the thermodynamical quantities as well as stability
globally and locally of above mentioned both BHs, respectively. In
the end of the paper, we summarized our results.

\section{Thermal Fluctuations}

In BH thermodynamics, the quantum fluctuations give rise to many
important problems and thermal fluctuations in the geometry of BH is
one of them. To solve this problem, it is necessary to contribute
the correction terms of entropy when the size of BH is reduced due
to the Hawking radiation and its temperature is increased. One can
neglect the correction terms for large BHs, as the thermal
fluctuations may not occur in it. Hence, the thermodynamics of BH is
modified by the thermal fluctuations and becomes more important for
smaller size BHs with sufficiently high temperature \cite{16a}.
Here, we analyze the effect of thermal fluctuations on the entropy
of general spherical symmetric metric, which can be done by
utilizing the Euclidean quantum gravity formalism
whose partition function can be defined as \cite{41,42,43,45,46}
\begin{equation}\label{1}
Y=\int DgDA e^{-I},
\end{equation}
where the Euclidean action is represented by  $I\rightarrow -iI$ for
this system. The partition function can be related to statistical
mechanical terms as follows \cite{47,48}
\begin{equation}\label{2}
Y=\int_{0}^{\infty}  DE\xi (E) e^{(-\gamma E)},
\end{equation}
where $\gamma = T^{-1}$.

Moreover, the density of states can be obtained with the help of
partition functions together with Laplace inverse as
\begin{equation}\label{3}
\xi(E)=\frac{1}{2 \pi i}\int_{\gamma_0-i \infty}^{\gamma_0+i \infty}
d\gamma e^{S(\gamma)},
\end{equation}
where $S=\gamma E+ \ln Y$. This entropy can be calculated by
neglecting all thermal fluctuations around the equilibrium
temperature $\gamma_0$. However, by utilizing the thermal
fluctuations along with Taylor series expansion around $\gamma_0$,
$S(\gamma)$ can be written as \cite{4,16a}
\begin{equation}\label{4}
    S=S_0+\frac{1}{2!}\big(\gamma-\gamma_0\big)^2\Big(\frac{\partial^2
    S(\gamma)}{\partial
    \gamma^2}\Big)_{\gamma=\gamma_0}+\frac{1}{3!}\big(\gamma-\gamma_0\big)^3\Big(\frac{\partial^3
    S(\gamma)}{\partial \gamma^3}\Big)_{\gamma=\gamma_0}+....
\end{equation}
As we know that the first derivative will vanish and hence density
of states turn out to be
\begin{equation}\label{5}
    \xi(E)=\frac{1}{2 \pi i}\int_{\gamma_0-i \infty}^{\gamma_0+i \infty}
    d\gamma e^{\frac{1}{2!}\big(\gamma-\gamma_0\big)^2\Big(\frac{\partial^2
    S(\gamma)}{\partial
    \gamma^2}\Big)_{\gamma=\gamma_0}+\frac{1}{3!}\big(\gamma-\gamma_0\big)^3\Big(\frac{\partial^3
    S(\gamma)}{\partial \gamma^3}\Big)_{\gamma=\gamma_0}}.
\end{equation}
Furthermore, by following \cite{4}, one can obtain the corrected
entropy
\begin{equation}\label{6}
S = S_0 - \frac{b}{2} \ln S_0T^2+\frac{c}{S_0},
\end{equation}
where $b$ and $c$ are introduced as constant parameters.
\begin{itemize}
  \item The original results can be obtained by setting $b,~c\rightarrow 0$,
i.e., the entropy without any correction terms. One can consider
this case for large BHs where temperature is very small.
  \item The usual logarithmic corrections can be recovered by setting
  $b
\rightarrow 1$ and $c \rightarrow 0$.
  \item The second order correction term  can be obtained by setting
  $b
\rightarrow 0$ and $c \rightarrow 1$ which represents the inverse
proportionality of original entropy.

  \item Finally, higher order corrections can be found by setting $b
\rightarrow 1$ and $c \rightarrow 1$.
\end{itemize}
Hence, the first order correction term represent the logarithmic
correction but the second order correction term represents the
inverse proportionality of original entropy. So, quantum correction
can be considered by these correction terms. As mentioned above, one
can avoid these correction terms for larger BHs. However, these
correction terms can be considered for BHs whose size decreases due
to Hawking radiation, while temperature increases and also thermal
fluctuation in the geometry of BH increases \cite{16a}.

\section{Thermodynamical Analysis of RN-AdS Black Hole with GM}

We consider the spherical symmetric metric of the form
\begin{equation}
ds^{2}_e=f(r)dt^{2}-(f(r))^{-1}dr^{2}-r^2({d\theta}^{2}
+\sin\theta{d\phi}^{2}),\label{1}
\end{equation}
where
\begin{equation}\label{7}
f(r)=1+2\big(\alpha - \frac{M}{r}\big)+\frac{Q^2}{r^2}-\frac{\Lambda
r^2}{3}.
\end{equation}
Here, $\alpha=\frac{\lambda^2}{2}$ is the GM charge, $\lambda$ is
the scale of gauge symmetric breaking $\lambda \sim 10^6$ GeV
\cite{17} and $M$ is the mass of BH. The metric function (\ref{7})
reduces to RN-AdS BH for $\alpha = 0$ while it becomes RN BH for
$\alpha=\Lambda =0$. By setting $f(r)=0$, which leads to
\begin{equation}\label{8}
r_+^4-\frac{3(1+2\alpha)}{\Lambda}r_+^2+\frac{6M}{\Lambda}r_+-\frac{3Q^2}{\Lambda}=0,
~~~~~\Lambda\neq 0,
\end{equation}
now set $d_1=-\frac{3(1+2\alpha)}{\Lambda}, d_2=+\frac{6M}{\Lambda},
d_3=-\frac{3Q^2}{\Lambda}$, we obtain the following algebraic
equation
\begin{equation}\label{8aa}
r_+^4+d_1 r_+^2+d_2 r_++d_3=0.
\end{equation}
By factorizing the above equation $X^2-Y^2=(X-Y)(X+Y)$ and solving
it, we obtain the resolvent cubic equation. The quantities $X$ and
$Y$ in perfect square are given by
\begin{equation}\label{8b}
    X=
    r^2+\frac{x}{2},~~~~~~~~~~~~~~~~Y=\sqrt{x-d_1}\Big(r-\frac{d_2}{2(x-d_1)}\Big),
\end{equation}
if the variable $x$ is chosen such that
\begin{equation}\label{8c}
x^3-d_1x^2-4d_3x+d=0,
\end{equation}
where $d=4d_1d_3-d_2^2=(36Q^2+72\alpha Q^2-36M^2)/\Lambda^2$ is the
resolvent cubic. Let $x_1$ be real roots of (\ref{8c}), the four
roots of original quadratic equation (\ref{8aa}) could be obtained
by following quadratic equation
\begin{equation}\label{8d}
    r_+^2\pm \sqrt{x_1-d_1}r_++\frac{1}{2}\Big(x_1\mp
    \frac{d_2}{\sqrt{x_1-d_1}}\Big),
\end{equation}
which are
\begin{eqnarray}
  r_{1+} &=& \frac{1}{2}\Big(\sqrt{x_1-d_1}+\sqrt{\Omega_-}\Big), \\
  r_{2+} &=& \frac{1}{2}\Big(\sqrt{x_1-d_1}-\sqrt{\Omega_-}\Big), \\
  r_{3+} &=& \frac{1}{2}\Big(\sqrt{x_1-d_1}+\sqrt{\Omega_+}\Big), \\
  r_{4+} &=&
  \frac{1}{2}\Big(\sqrt{x_1-d_1}-\sqrt{\Omega_+}\Big).
\end{eqnarray}
where $\Omega_{\pm}=-(x_1+d_1)\pm \frac{2d_2}{\sqrt{x_1-d_1}}$ and
$x_1>d_1=-\frac{3(1+2\alpha)}{\Lambda}$. The polynomial (\ref{8})
has at most three real roots, which are Cauchy, cosmological and
event horizons \cite{17}.

The mass, entropy, volume and temperature of RN-AdS BH with GM in
horizon radius form can be written as
\begin{eqnarray}\label{9}
    M|_{r=r_+}&=&\frac{-r_+^4 \Lambda+6\alpha r_+^2+3Q^2+3 r_+^2}{6r_+},\\\label{10}
    S_0|_{r=r_+}&=&\pi r_+^2,\\\label{11}
    V|_{r=r_+}&=&\frac{4}{3}\pi r_+^3,\\\label{tm1}
    T|_{r=r_+} &=& \frac{f'(r)}{4 \pi}=\frac{-r_+^4 \Lambda+3M r_+-3Q^2}{6 \pi r_+^3},
\end{eqnarray}
where $r_+\neq0$. We can analyze the thermodynamics of RN-AdS BH
with GM in terms of mass $M$, horizon radius $r_+$, cosmological
constant $\Lambda$ and charge $Q$. Inserting the mass $M$ in above
equation, the temperature reduces to
\begin{equation}\label{12t}
    T|_{r=r_+}=\frac{-r_+^4\Lambda +2 \alpha r_+^2-Q^2+r_+^2}{4\pi r_+^3}.
\end{equation}
It is clear that the temperature is decreasing function of horizon
radius, so when the size of black hole decreased, the temperature
grow up and thermal fluctuations will be important as mentioned
before. For real positive temperature, we have the following
condition
\begin{equation}\label{13}
r_+^2 \geq \frac{(2 \alpha+1)\pm \sqrt{(2 \alpha+1)^2-4 \Lambda
Q^2}}{2\Lambda}\quad ~~~\text{with}~~~(2 \alpha+1)^2\geq 4 \Lambda
Q^2.
\end{equation}
The corrected entropy of RN-AdS BH with GM can be obtained by using
Eqs.(\ref{6}), (\ref{10}) and (\ref{12t}), which turns out to be
\begin{equation}\label{15s}
S|_{r=r_+}=\pi r_+^2-\frac{b}{2} \ln\Big(\frac{(-r_+^4 \Lambda +2
\alpha r_+^2-Q^2+r_+^2)^2}{16 \pi r_+^4}\Big)+\frac{c}{\pi r_+^2}.
\end{equation}
The pressure can also be calculated in view of Eqs.(\ref{11}),
(\ref{12t}), (\ref{15s}) as
\begin{eqnarray}\nonumber
P|_{r=r_+}&=&T\Big(\frac{\partial S}{\partial V}\Big)_V=\frac{1}{8
\pi^3 r_+^8}\Big(-\pi^2 r_+^8 \Lambda+\pi(2\pi \alpha
+b\Lambda+\pi)r_+^6+(c\Lambda-\pi^2 Q^2)r_+^4\\\label{16p}&-&(\pi b
Q^2+2\alpha c+c)r_+^2+cQ^2\Big).
\end{eqnarray}
\begin{figure}
\centering
\begin{tabular}{@{}cccc@{}}
\includegraphics[width=.5\textwidth]{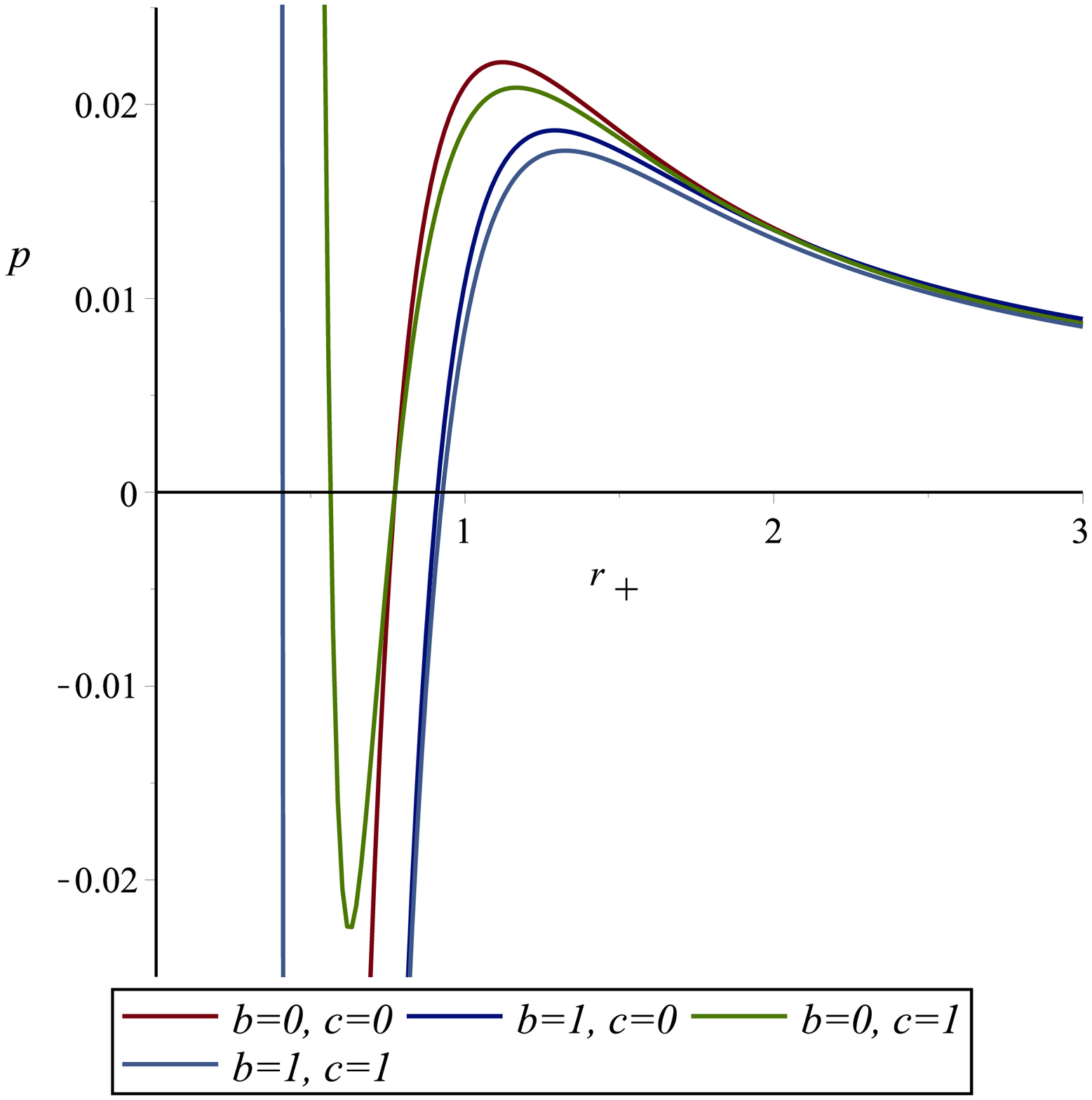} &
\includegraphics[width=.5\textwidth]{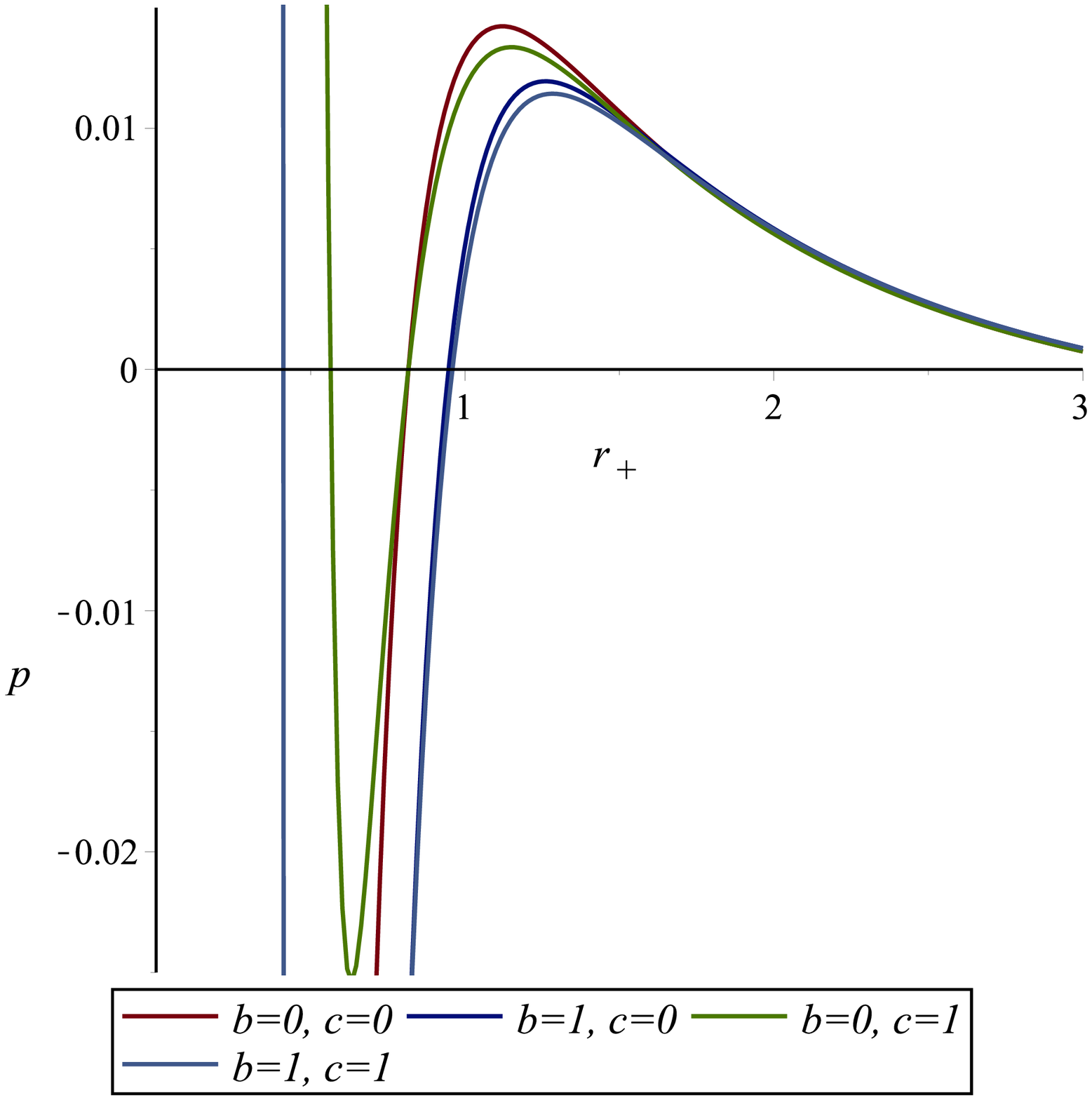}
\end{tabular}
\caption{The plot of pressure for RN-AdS BH with GM for $\alpha =
0.075$, $Q=0.85$ and $\Lambda = -0.1$ (left panel), $\Lambda = 0.1$
(right panel).}
\end{figure}

In Fig. \textbf{1}, we discuss the behavior of pressure for negative
and positive values of cosmological constant. In both panels, we
observe that the pressure decreases due to correction terms. For
$b=1, c=1$, we observe the lowest trajectory of pressure then above
it we have the trajectory of pressure at $b=1,c=0$ which becomes the
logarithmic correction term. Furthermore, we see that the pressure
increases for $b=0, c=1$ which is the second order correction term.
The pressure is maximum if we avoid these correction terms. Also, we
observe in both panels that the pressure increases for negative
cosmological constant. However, the pressure becomes zero at
$r_+=3.3$ for positive value of $\Lambda$ and becomes negative for
large horizon while remains positive for negative value of
$\Lambda$. Hence, we can conclude that the pressure decreases due to
higher order correction terms and positive values of $\Lambda$. BHs
derived from general relativity coupled with matter fields with
$P\leq 0$ are thermodynamically unstable.

\subsection{Stability of RN-AdS BH with GM}

Here, we shall discuss the thermodynamical stability of RN-AdS with
GM for which we can define
\begin{equation}\label{17e}
E|_{r=r_+} = \int T dS.
\end{equation}
In BH thermodynamics, an important measurable physical quantity is
the heat capacity or thermal capacity. The heat capacity of the BH
may be stable or unstable by observing its sign (positive or
negative), respectively. There are two types of heat capacities
corresponding to a system such as $C_v$ and $C_p$ which determine
the specific heat with constant volume and pressure, respectively.
$C_v$ can be defined as
\begin{equation}\label{18}
C_V|_{r=r_+} = T \left(\frac{\partial S}{\partial T}\right)_V.
\end{equation}
Eqs. (\ref{12t}) and (\ref{15s}) lead to
\begin{eqnarray}\nonumber
C_V|_{r=r_+}&=&-\frac{2}{\pi r_+^2 (r_+^4 \Lambda +2\alpha
r_+^2-3Q^2+r_+^2)} \Big(-\pi^2r_+^8 \Lambda +2\pi^2 \alpha r_+^6+\pi
br_+^6
\Lambda\\
&-&\pi^2 Q^2r_+^4+\pi^2 r^6-\pi b Q^2 r_+^2+cr_+^4\Lambda - 2\alpha
c r_+^2+cQ^2-cr_+^2\Big).\label{18a}
\end{eqnarray}
Moreover, the $C_p$ can be evaluated as
\begin{equation}\label{17}
 C_P|_{r=r_+} = \left(\frac{\partial(E + PV)}{\partial T}\right)_P,
\end{equation}
and its expression can be obtained with the help of Eqs. (\ref{11}),
(\ref{12t}), (\ref{16p}) and (\ref{17e}) as
\begin{eqnarray}\nonumber
C_P|_{r=r_+} &=& -\frac{4}{\pi r_+^2 (r_+^4 \Lambda +2\alpha
r_+^2-3Q^2+r_+^2)}\Big(-3\pi^2r_+^8\Lambda
+4\pi^2\alpha r_+^6 \\
&+&2\pi b r_+^6 \Lambda-\pi^2 Q^2r_+^4+2\pi^2r_+^6
-cr_+^4\Lambda-cQ^2\Big).\label{17a}
\end{eqnarray}
\begin{figure}
\centering
  \begin{tabular}{@{}cccc@{}}
    \includegraphics[width=.5\textwidth]{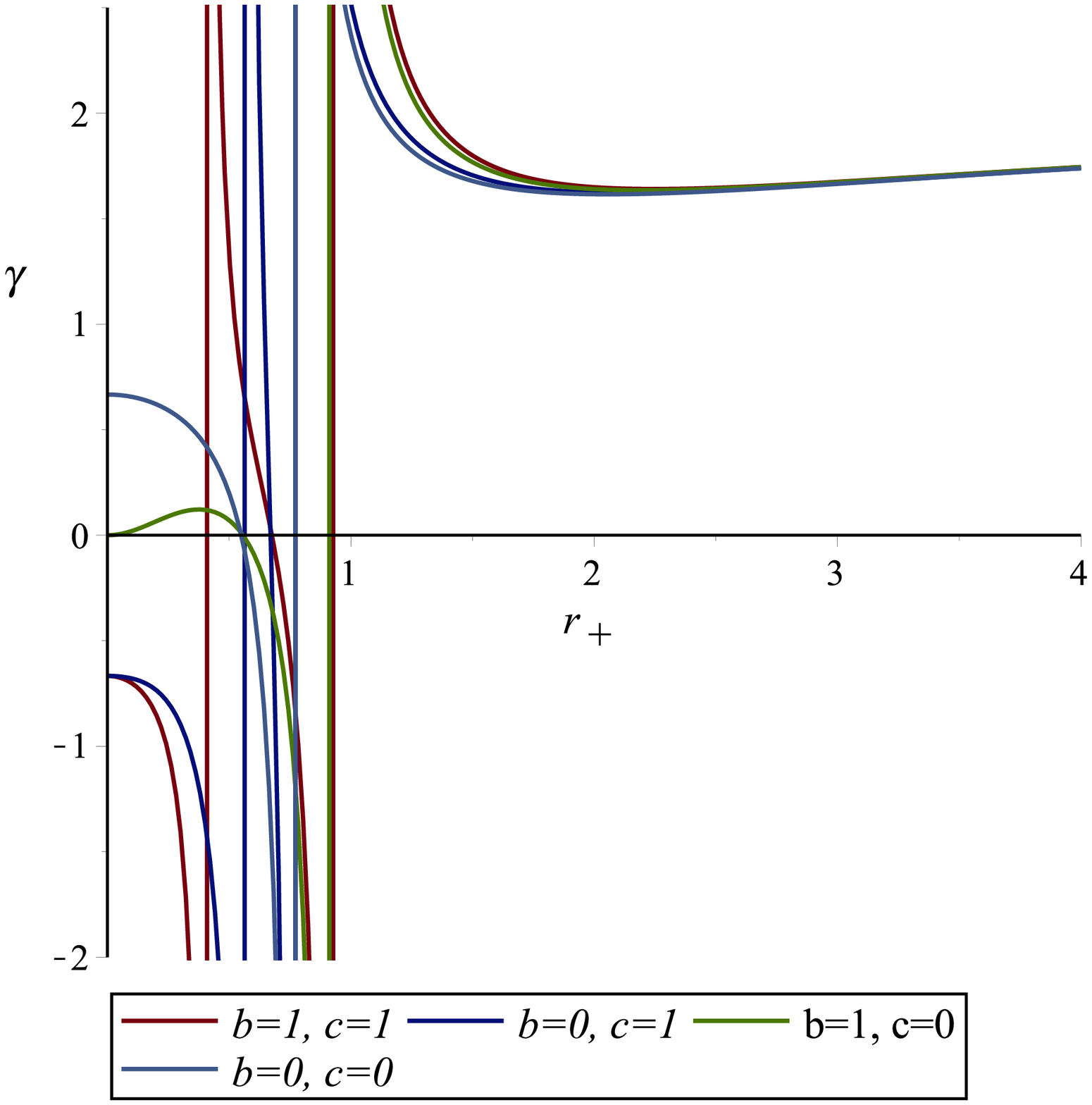} &
    \includegraphics[width=.5\textwidth]{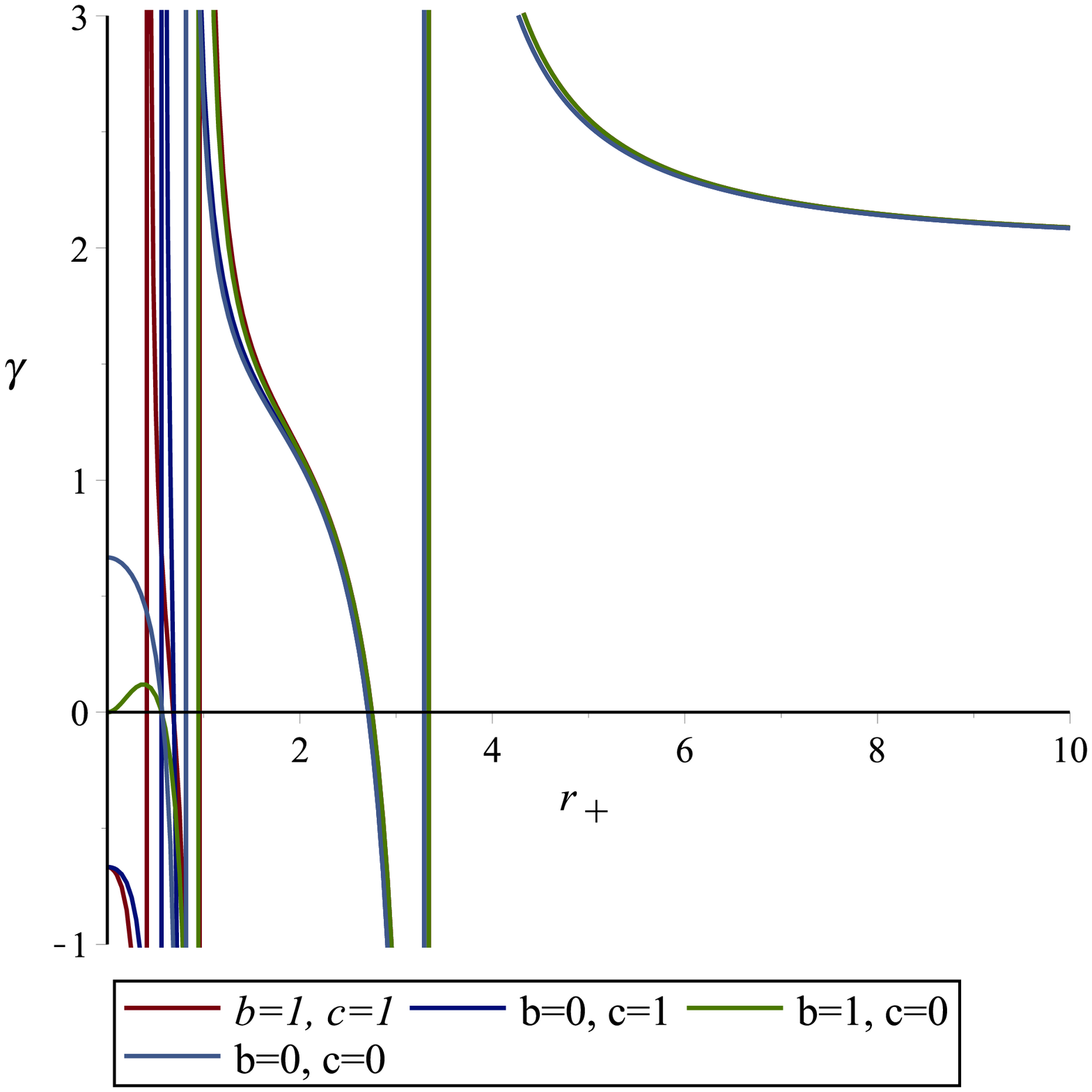}
    \end{tabular}
\caption{The plot of $\gamma$ versus horizon radius for RN-AdS BH
with GM for $\alpha = 0.075$, $Q=0.85$ and $\Lambda = -0.1$ (left
panel), $\Lambda = 0.1$ (right panel).}
\end{figure}
The above two specific heat relations can be comprises into a ratio
that is denoted by $\gamma = C_p/C_v$ and its plot is given in Fig.
\textbf{2}. It can be noted that the value of $\gamma$ increases due
to the correction terms. We find maximum value of $\gamma$ as
$\gamma \rightarrow 2.1$ for positive value of $\Lambda$ and larger
horizon, while for negative value of cosmological constant $\gamma
\rightarrow 1.8$. Thus, one can observe that $\gamma$ exhibits more
stable behavior for lower value of $\Lambda$ as compare to positive
value of $\Lambda$. In left panel of Fig. \textbf{2}, we observe
that the value of $\gamma$ becomes higher by utilizing both
correction and logarithmic correction terms in small horizon. Hence,
it is pointed out that the value of $\gamma$ becomes higher due to
correction terms and exhibits more stable behavior for lower values
of $\Lambda$.

\subsection{Phase transition}

The stability of BH can be analyzed through $C_v$ (Eq. (\ref{18a}))
because $C_v\lesseqgtr0$ corresponds to local stability of BH, phase
transition and local instability of the BH.
\begin{table}[h]
\begin{center}
\begin{tabular}{|c|c|c|c|}
\hline
$\Lambda$&\begin{tabular}{@{}c@{}c@{}}correction\\
terms\end{tabular}&\begin{tabular}{@{}c@{}c@{}} range of \\local
stability\end{tabular}&phase transition\\
 \hline $-0.1$ &
\begin{tabular}{@{}c@{}c@{}c@{}}$b=1$, $c=1$ \\ $b=0$,
$c=1$\\ $b=1$, $c=0$\\$b=0$,
$c=0$\end{tabular}&\begin{tabular}{@{}c@{}c@{}c@{}}
$0<r_+<0.4,0.92<r_+<1.53,r_+>3$\\ $0<r_+<0.55,0.76<r_+<1.53,~r_+>3$\\
$0.91<r_+<1.53,~r_+>3$\\$0.77<r_+<1.53,~r_+>3$\end{tabular}&
\begin{tabular}{@{}c@{}c@{}c@{}c@{}}$0.41,~0.93$\\$0.56,~0.77$\\$0.91$\\$0.77$\end{tabular}\\
\hline 0.1&\begin{tabular}{@{}c@{}c@{}c@{}c@{}}$b=1$, $c=1$
\\
$b=0$, $c=1$
\\$b=1$, $c=0$\\$b=0$, $c=0$\end{tabular}&\begin{tabular}{@{}c@{}c@{}c@{}c@{}}
$0<r_+<0.41,~0.95<r_+<1.27,~r_+>3.34$ \\ $0<r_+<0.56,~0.81<r_+<1.27,~r_+>3.29$\\
$0.95<r_+<1.27,~r_+>3.34$\\$0.81<r_+<1.27,~r_+>3.33$\end{tabular}&\begin{tabular}
{@{}c@{}c@{}c@{}c@{}}$0.41,~0.96,~3.34$\\$0.56,~0.82,~3.29$\\$0.95,~3.34$\\$0.82,~3.29$\end{tabular}\\
\hline
\end{tabular}
\end{center}
\caption{Range of local stability and critical points of horizon
radius of phase transition for RN-AdS BH with GM due to the effect
of higher order correction entropy.}\label{t1}
\end{table}
\begin{figure} \centering
  \begin{tabular}{@{}cccc@{}}
    \includegraphics[width=.5\textwidth]{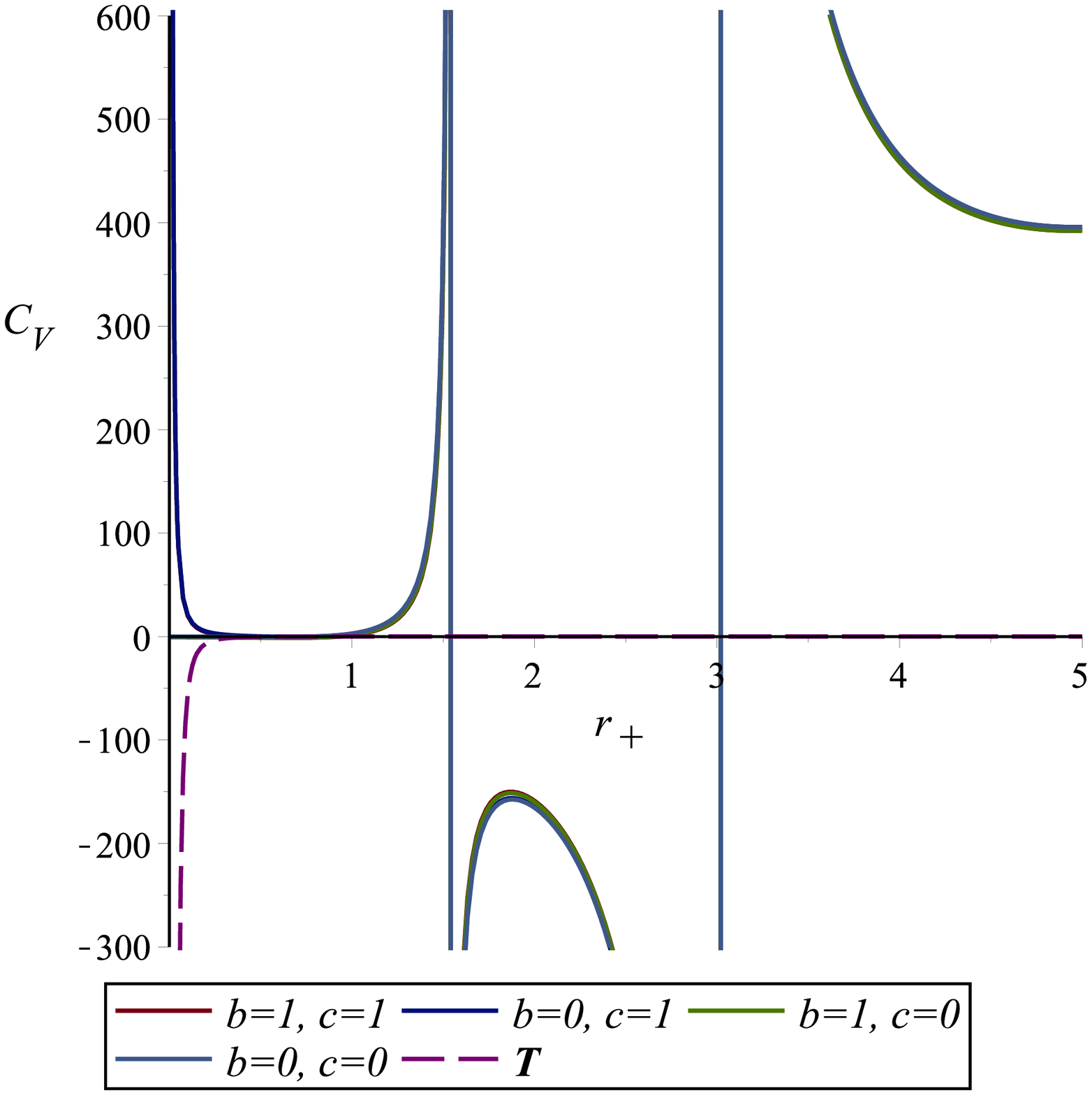} &
    \includegraphics[width=.5\textwidth]{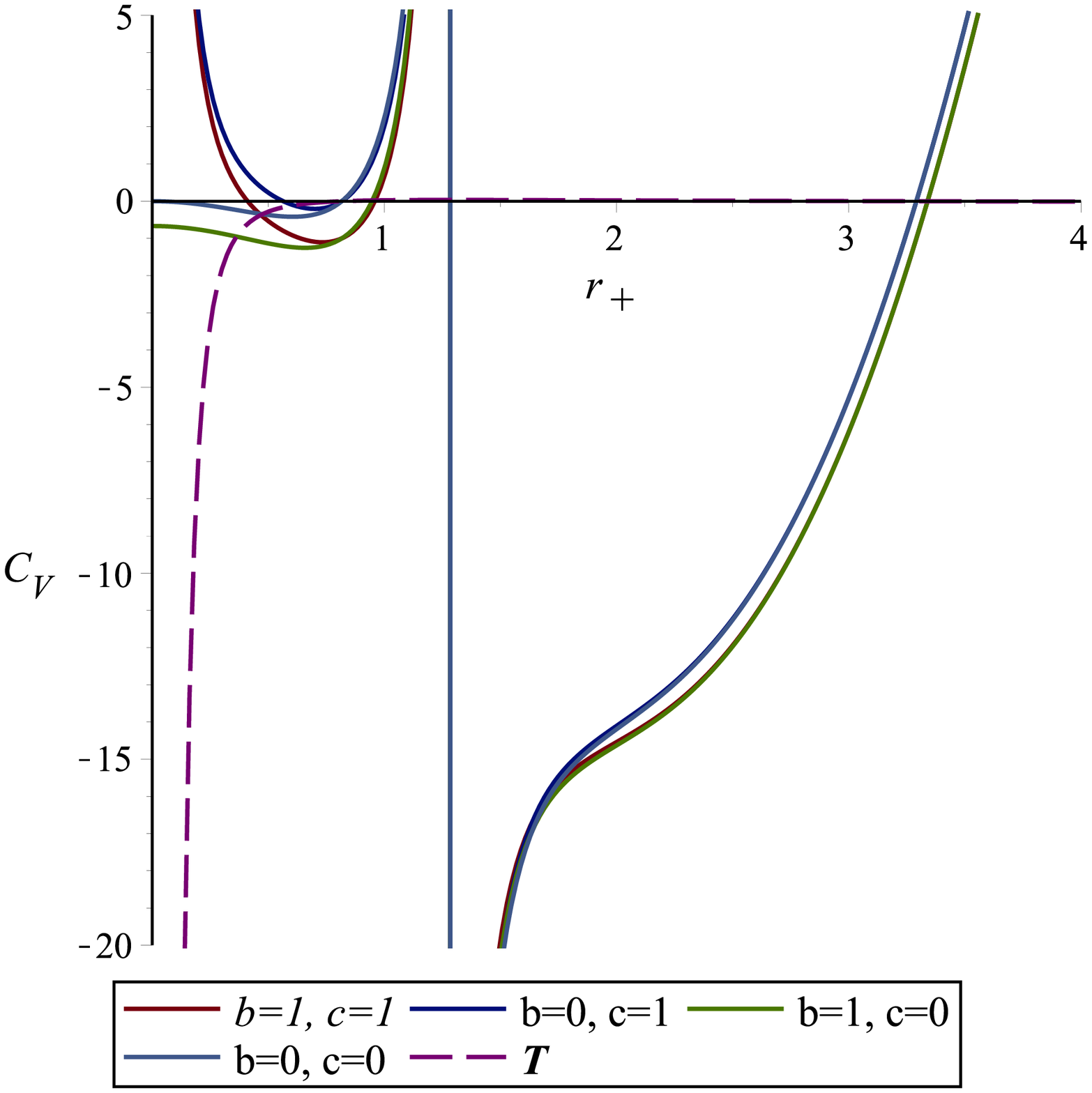}
    \end{tabular}
\caption{The plot of specific heat at constant volume for RN-AdS BH
with GM for $\alpha = 0.075$, $Q=0.85$ and $\Lambda = -0.1$ (left
panel), $\Lambda = 0.1$ (right panel) }
  \end{figure}

We find the range of BH horizon of locally thermodynamical stability
due to thermal fluctuations for negative and positive values of
cosmological constant in left and right panels of Fig. \textbf{3},
respectively and in Table \textbf{1}. We can observe from Table
\textbf{1} that the range of local stability is maximum for second
order correction terms ($b=0, c=1$) as well as for both correction
terms ($b=1,c=1$) as compare to the others in both cases of
$\Lambda$. Moreover, we obtain the phase transition points for both
case in Table \textbf{\ref{t1}}. We observe that one can obtain more
phase transition points for $b=0,c=1$ and $b=1,c=1$ as compare to
others in both cases. Also, we find more phase transition points for
positive value of cosmological constant as compare to negative
value. It is noted that RN-AdS BH with GM is the most locally stable
by considering the both correction terms for negative cosmological
constant. We also find the different horizon regions where the BH is
stable. For instance, it is completely stable for $r_+ > 3$ in the
case of negative $\Lambda$ while for positive $\Lambda$, it is
completely stable for $r_+ \geq 3$. Hence, it can be concluded that
if we increases the value of $\Lambda$ and consider the second order
or both correction terms then we can find more phase transition
points and the range of local stability is also increased.

\subsection{Grand Canonical Ensemble}

The BH can be considered as a thermodynamical object by treating it
as grand canonical ensemble system where fix chemical charge is
represented by $\mu = \frac{Q}{r_+}$. The temperature increases due
to effect of chemical potential. The grand canonical ensemble is
also known as Gibb's free energy, which can be defined as
\begin{equation}\label{c21}
G = M - T S-\mu Q,
\end{equation}
and it turns out to be
\begin{eqnarray}\nonumber
  G|_{r=r_+} &=& -\frac{-r_+^4\Lambda+(2\alpha+1)r_+^2-Q^2}{8 \pi^2 r_+^5}\Big(-2\pi^2r_+^4-2c\\
  &+&b\ln \big(\frac{(-r_+^4\Lambda+2\alpha r_+^2-Q^2+r_+^2)^2}{16 \pi
  r_+^4}\big)\Big)-\mu r_+.\label{g1}
\end{eqnarray}

\begin{figure}
\centering
  \begin{tabular}{@{}cccc@{}}
    \includegraphics[width=.5\textwidth]{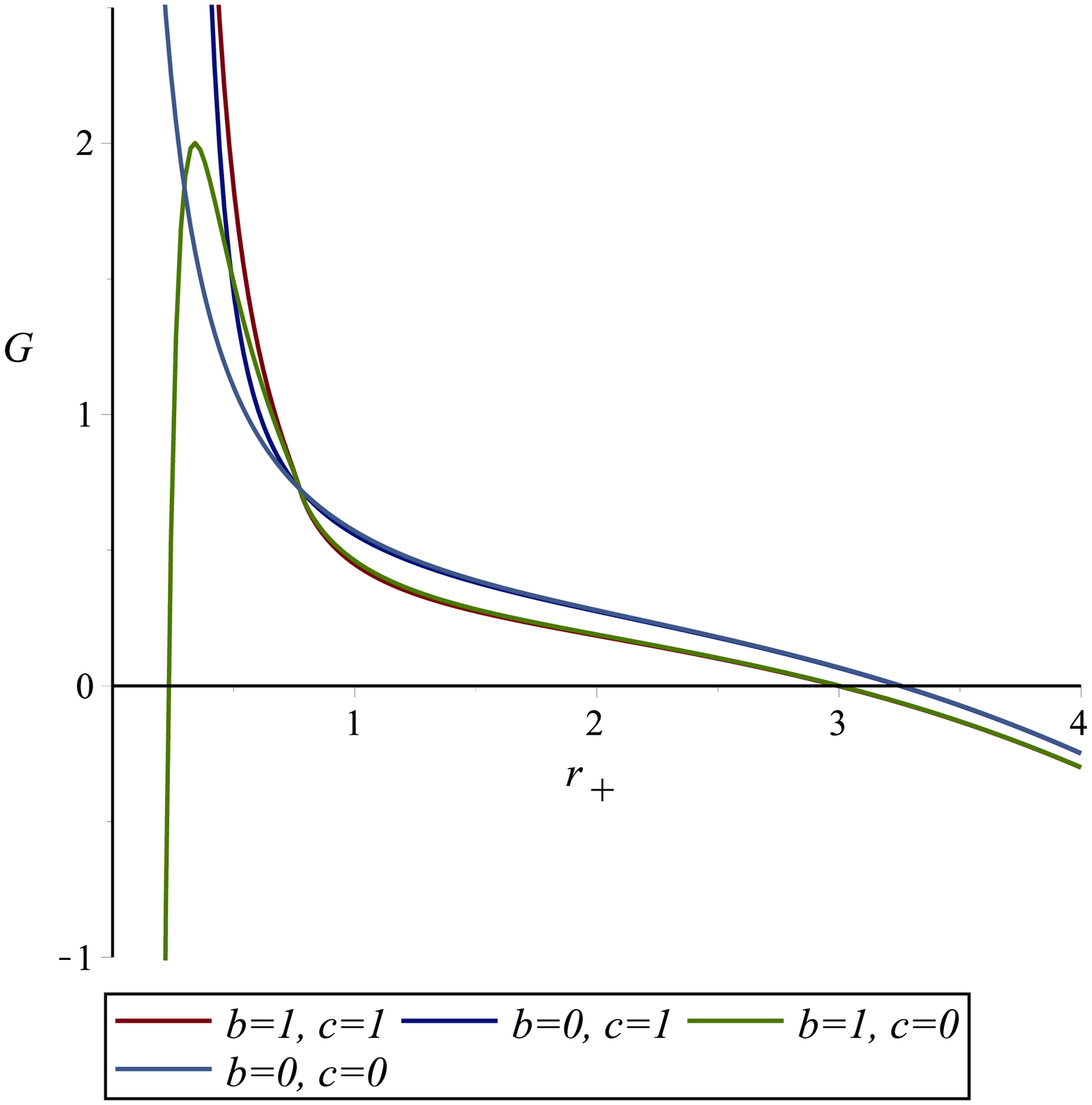} &
    \includegraphics[width=.5\textwidth]{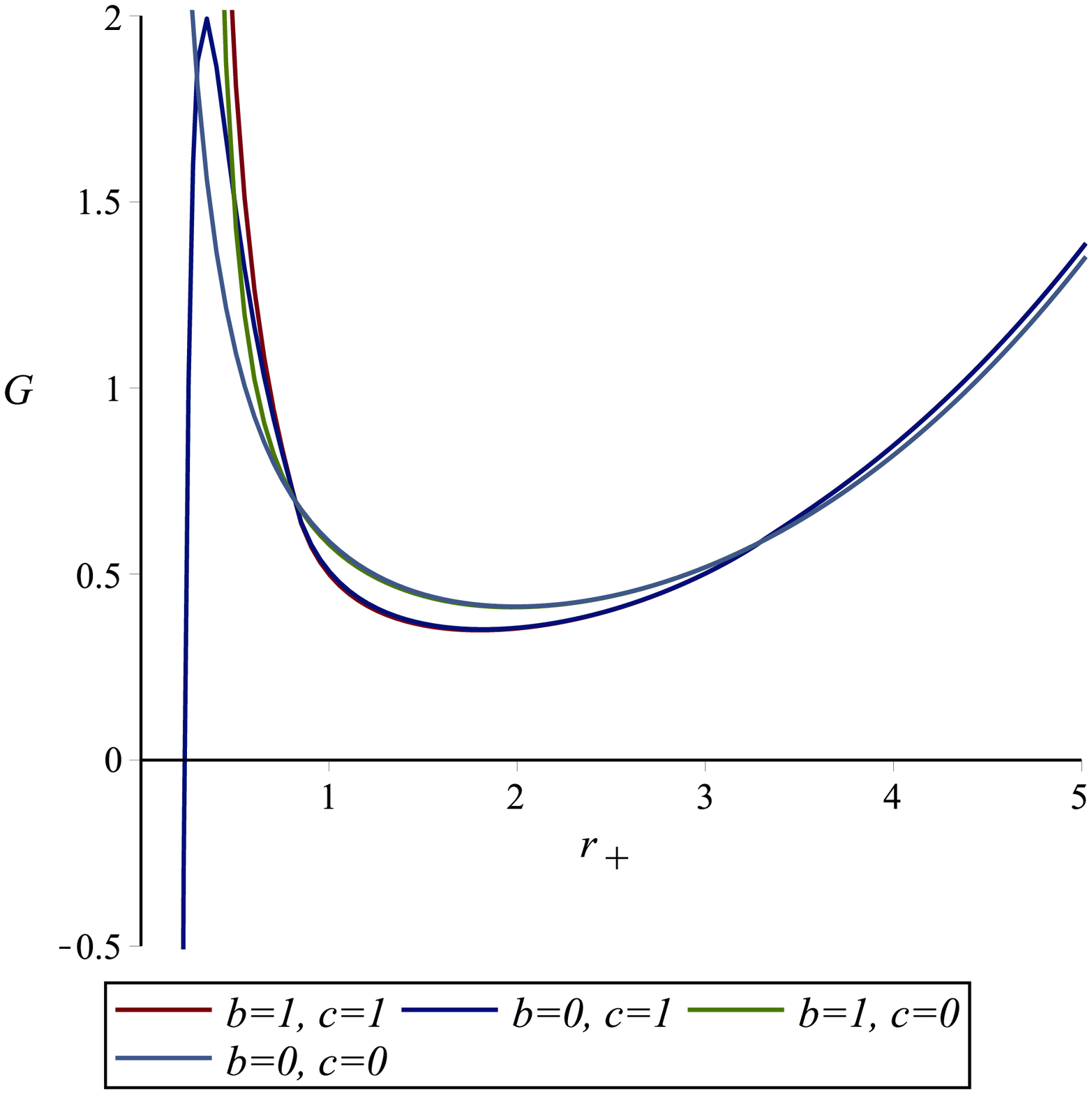}
    \end{tabular}
\caption{The plot of Gibb's free energy for RN-AdS BH with GM for
$\alpha = 0.075$, $Q=0.85$ and $\Lambda = -0.1$ (left panel),
$\Lambda = 0.1$ (right panel) }
  \end{figure}
Fig. \textbf{4} indicates the view of Gibb's free energy for
negative and positive cosmological constant. It can be noted from
both panels of this figure that the correction terms reduce the
Gibb's free energy. In left panel, we observe that the BH is most
stable for higher order correction terms ($b=1, c=1$ and $b=0, c=1$)
near the singularity while it exhibits the most stable behavior at
$r_+ > 0.8$ for second order correction term ($b=0, c=1$). In right
panel, it is noticed that the BH is most stable for both correction
terms ($b=1, c=1$) in the case for positive cosmological constant.
Hence, it is concluded that RN-AdS with GM is the most stable BH for
positive cosmological constant and higher order correction terms.

\subsection{Canonical Ensemble}

\begin{figure}
\centering
  \begin{tabular}{@{}cccc@{}}
    \includegraphics[width=.5\textwidth]{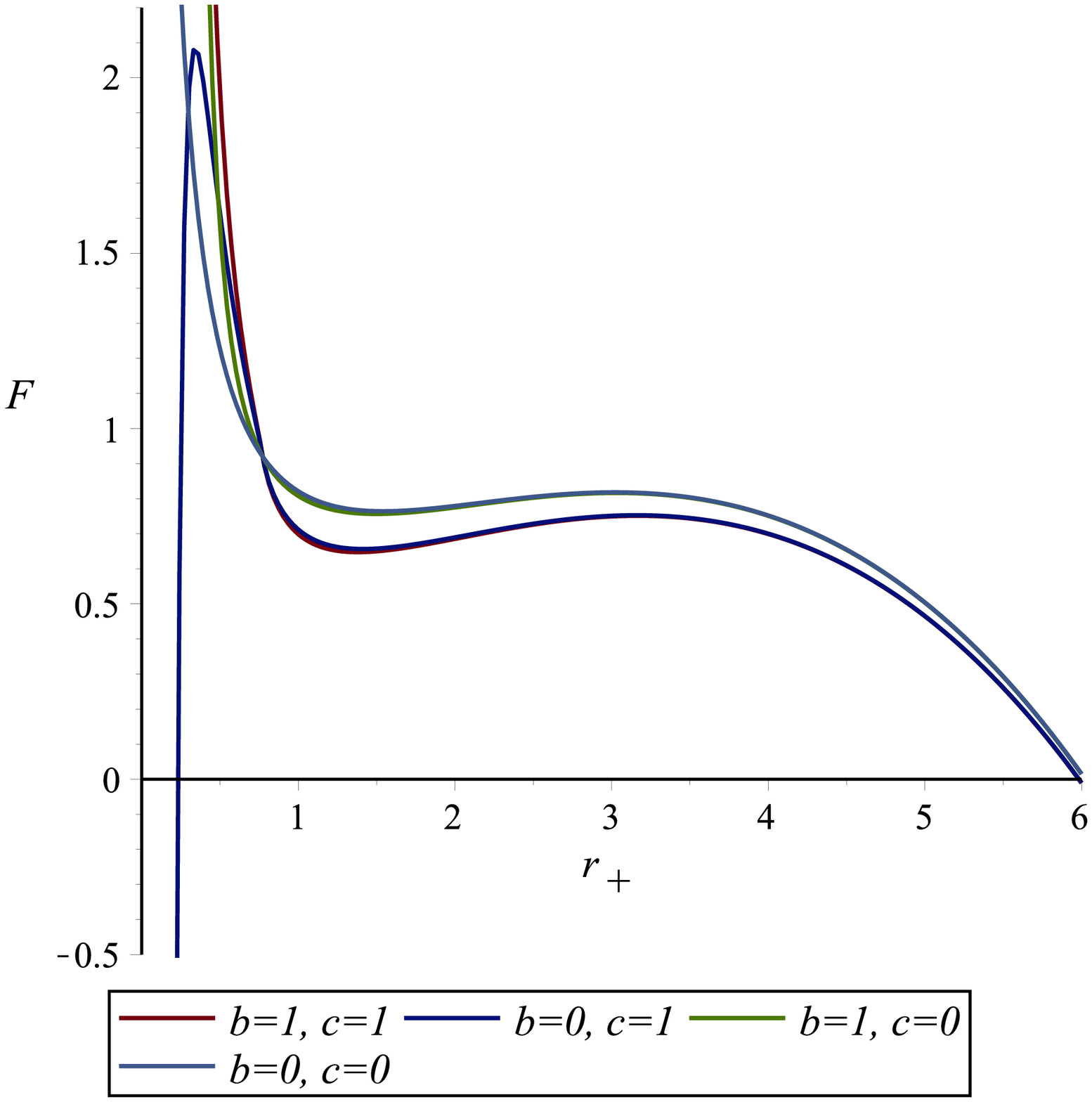} &
    \includegraphics[width=.5\textwidth]{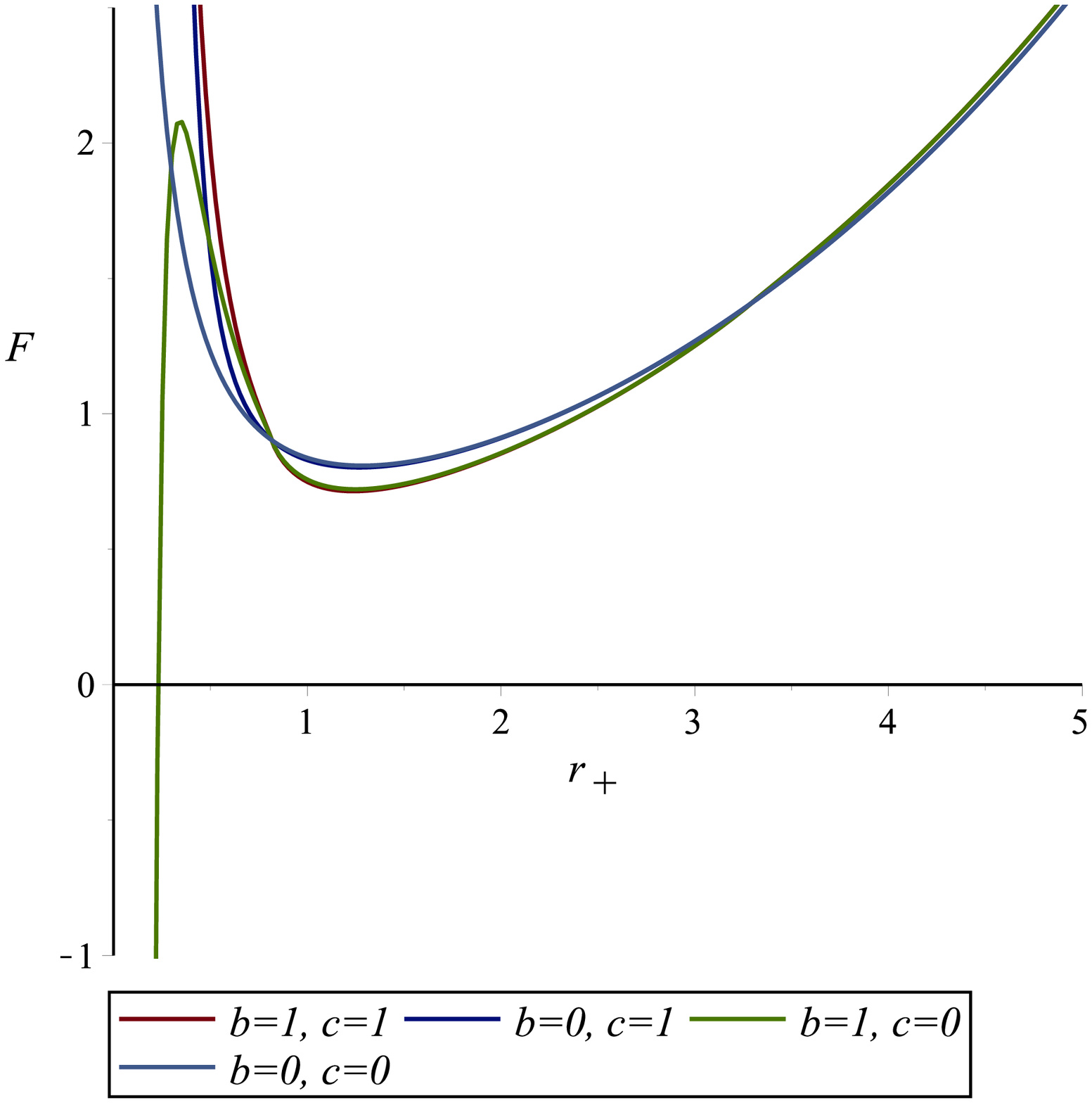}
    \end{tabular}
\caption{The plot of Helmhotz free energy versus horizon radius for
RN-AdS BH with GM for $\alpha = 0.075$, $Q=0.85$ and $\Lambda =
-0.1$ (left panel), $\Lambda = 0.1$ (right panel.}
  \end{figure}
Here, we consider BH as a canonical ensemble (closed system) where
transformation of charge is prohibited. For fixed charge, the free
energy in canonical ensemble form termed as Helmhotz free energy and
can be defined as
\begin{equation}\label{22}
F = M- TS,
\end{equation}
which becomes
\begin{eqnarray}\nonumber
F|_{r=r_+} &=& -\Big(-2\pi^2r_+^4-2c+ b\ln
\big(\frac{(-r_+^4\Lambda+2\alpha r_+^2-Q^2+r_+^2)^2}{16 \pi
r_+^4}\big)\Big)\\\label{f1}&\times&\frac{-r_+^4\Lambda+(2\alpha+1)r_+^2-Q^2}{8
\pi^2 r_+^5}.
\end{eqnarray}
The behavior of Helmhotz free energy for negative and positive
cosmological constant is plotted in Fig. \textbf{5}. In both panels,
we observe that Helmhotz free energy decreases due to correction
terms. For negative value of $\Lambda$, the BH is most stable for
both correction terms ($b=1, c=1$) near the singularity while at
$r_+>0.8$, it is the most stable for second order correction ($b=0,
c=1$). For positive value of $\Lambda$, BH exhibits the most stable
behavior for both correction terms at $r_+<0.8$ and $r_+>3.4$ while
it is the most stable for second order correction term for
$0.8<r_+<3.4$ approximately.

\section{Thermodynamical Analysis of $f(R)$ BH}

We consider the spherical symmetric metric of $f(R)$ BH of the form
\cite{49}
\begin{equation}
ds^{2}=f(r)dt^{2}-(f(r))^{-1}dr^{2}-r^2({d\theta}^{2}
+\sin\theta{d\phi}^{2}),\label{1}
\end{equation}
where
\begin{equation}\label{7a}
f(r)=1+\frac{2M}{r}+\beta r-\frac{\Lambda r^2}{3}.
\end{equation}
Here, $M$ is the mass of the BH, $\beta=a/d\geq 0$ is a constant
with $d$ is the scale factor and $a$ is the dimensionless parameter
\cite{49}. The horizon radius can be obtained through metric
function (\ref{7a}) as
\begin{equation}\label{8a}
    r_+^3 \Lambda--3r_+^2 \beta +6M-3r_+ = 0.
\end{equation}

One can obtain the mass and temperature of $f(R)$ BH at horizon
radius as
\begin{eqnarray}\label{9a}
    M|_{r=r_+}=\frac{-r_+^3 \Lambda+3\beta r_+^2+3 r_+}{6},\label{tm2}
  T|_{r=r_+} &=& \frac{-2r_+^3 \Lambda+3\beta r_+^2+6M}{12 \pi
   r_+^2}.
\end{eqnarray}
By utilizing the mass $M$ in above equation, the temperature reduces
to
\begin{equation}\label{12tm}
    T|_{r=r_+}=\frac{-r_+^2\Lambda +2 r_+ \beta + 1}{4\pi r_+}.
\end{equation}
For real and positivity of temperature, we have
\begin{equation}\label{13r}
r_+ \geq \frac{ \beta + \sqrt{\beta ^2 +\Lambda }}{ \Lambda}.
\end{equation}
The condition satisfied
\begin{equation}\label{14a}
    \beta ^2 +\Lambda \geq 0.
\end{equation}
Since, the temperature is decreasing function of horizon radius, so
when the size of black hole decreased, the temperature grow up and
thermal fluctuations will be important. The higher order corrected
entropy and pressure for this BH take the form
\begin{eqnarray}\label{15}
    S|_{r=r_+}&=&\pi r_+^2-\frac{b}{2} \ln(\frac{(r_+^2 \Lambda -2 \beta r_+
    -1)^2}{16 \pi})+\frac{c}{\pi r_+^2},\\\nonumber
    P|_{r=r_+}&=&\frac{1}{8 \pi^3 r_+^6}\big(-\pi^2r_+^6\Lambda+2\pi^2r_+^5\beta+b\pi
    r_+^4\Lambda- b\pi r_+^3\beta+\pi^2 r_+^4+cr_+^2 \Lambda\\\label{16a}
    &-&2cr_+
    \beta-c\big).
\end{eqnarray}
\begin{figure}
\centering
  \begin{tabular}{@{}cccc@{}}
    \includegraphics[width=.5\textwidth]{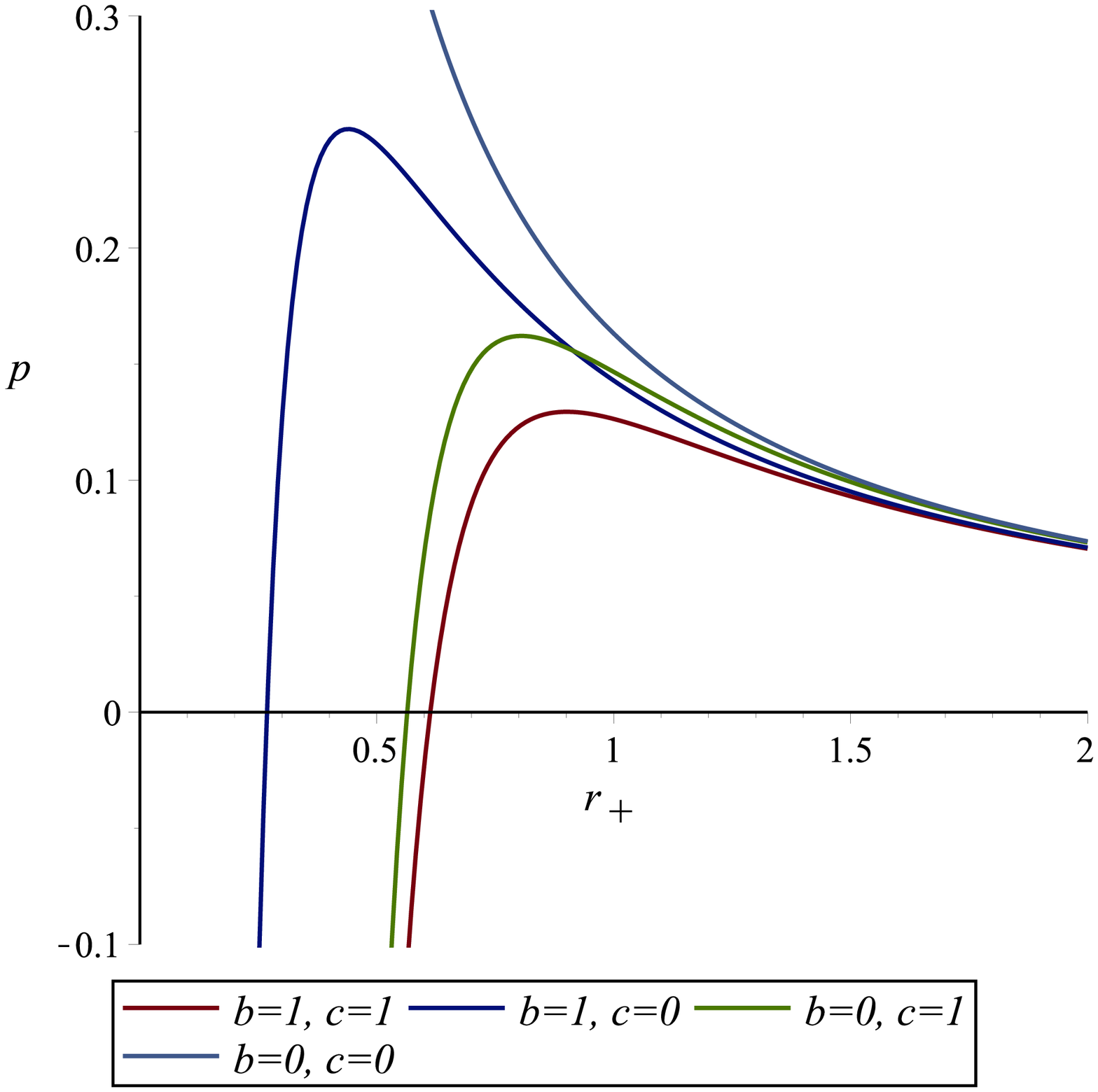} &
    \includegraphics[width=.5\textwidth]{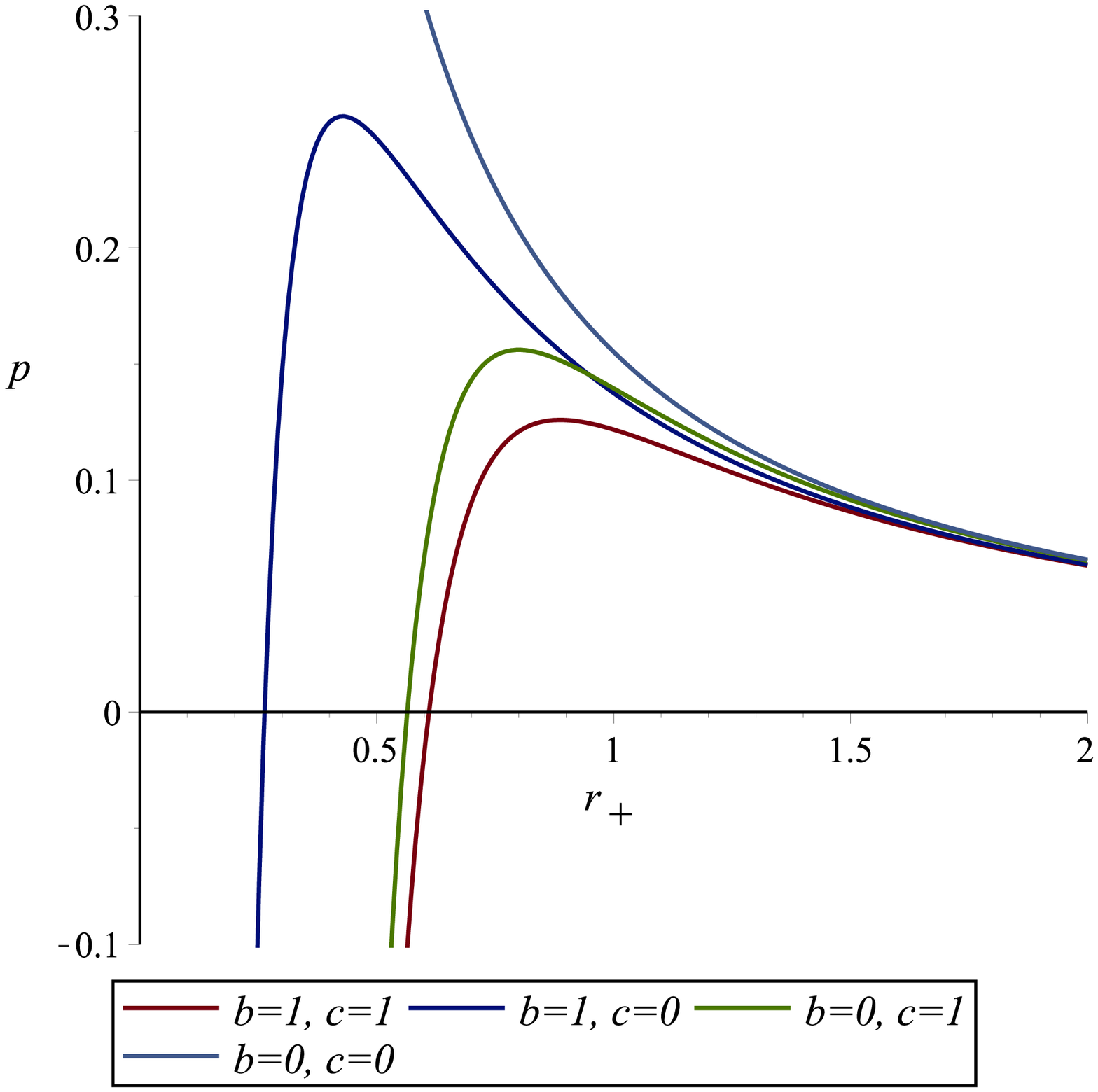}
    \end{tabular}
\caption{The plot of pressure for $f(R)$ BH for $\beta = 1.5$ and
$\Lambda = -0.1$ (left panel), $\Lambda = 0.1$ (right panel) }
  \end{figure}
In Fig. \textbf{6}, we analyze the trajectories of pressure for
negative and positive values of $\Lambda$. We can see pressure
decreases due to correction terms. We obtain highest pressure in the
absence of correction terms and the lowest pressure in the presence
of both correction terms $(b=1, c=1)$. Also, we observe that the
pressure is lower for second order correction term $(b=0, c=1)$ as
compare to logarithmic correction term $(b=1, c=0)$. Interestingly,
we find no change in pressure with respect to cosmological constant.
Hence, it can be concluded that the pressure decreases due to higher
order correction terms only.

\subsection{Stability of $f(R)$ BH}

For stability of $f(R)$ BH, we can obtain $C_v$ by using Eqs.
(\ref{12tm}) and (\ref{15}) as follows
\begin{eqnarray}\nonumber
C_V|_{r=r_+}&=&-\frac{2}{\pi r_+^2 (r_+^2 \Lambda +1)}
\Big(-\pi^2r_+^6 \Lambda +2\pi^2 \beta r_+^5+\pi br_+^4
\Lambda\\
&-&\pi r_+^3 \beta+\pi^2 r_+^4+cr_+^2\Lambda - 2\beta c
r_+-c\Big).\label{18fa}
\end{eqnarray}
Moreover, $C_p$ can be obtained by using Eqs. (\ref{11}),
(\ref{12tm}) and (\ref{16a}), we have
\begin{figure}
\centering
  \begin{tabular}{@{}cccc@{}}
    \includegraphics[width=.5\textwidth]{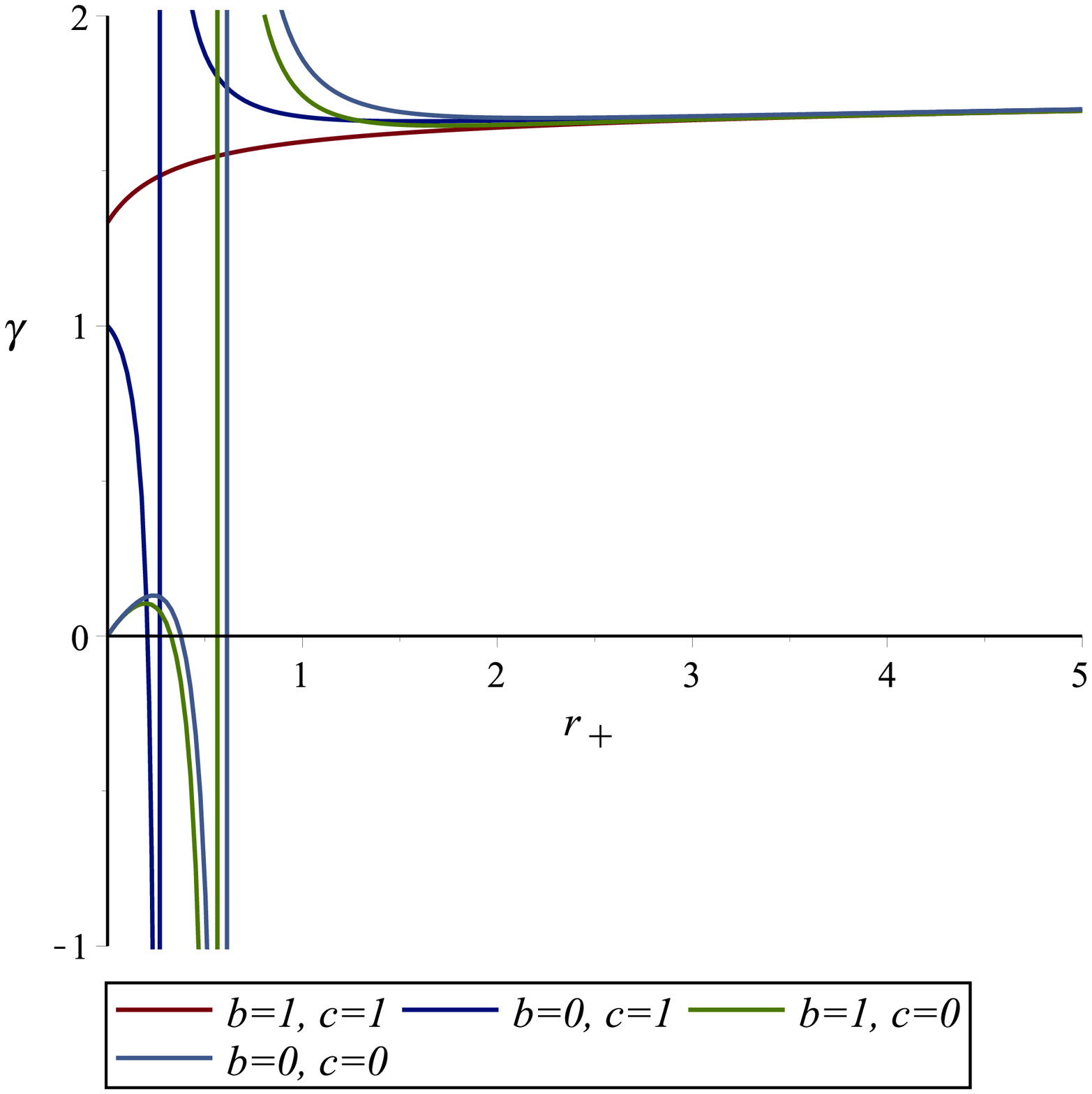} &
    \includegraphics[width=.5\textwidth]{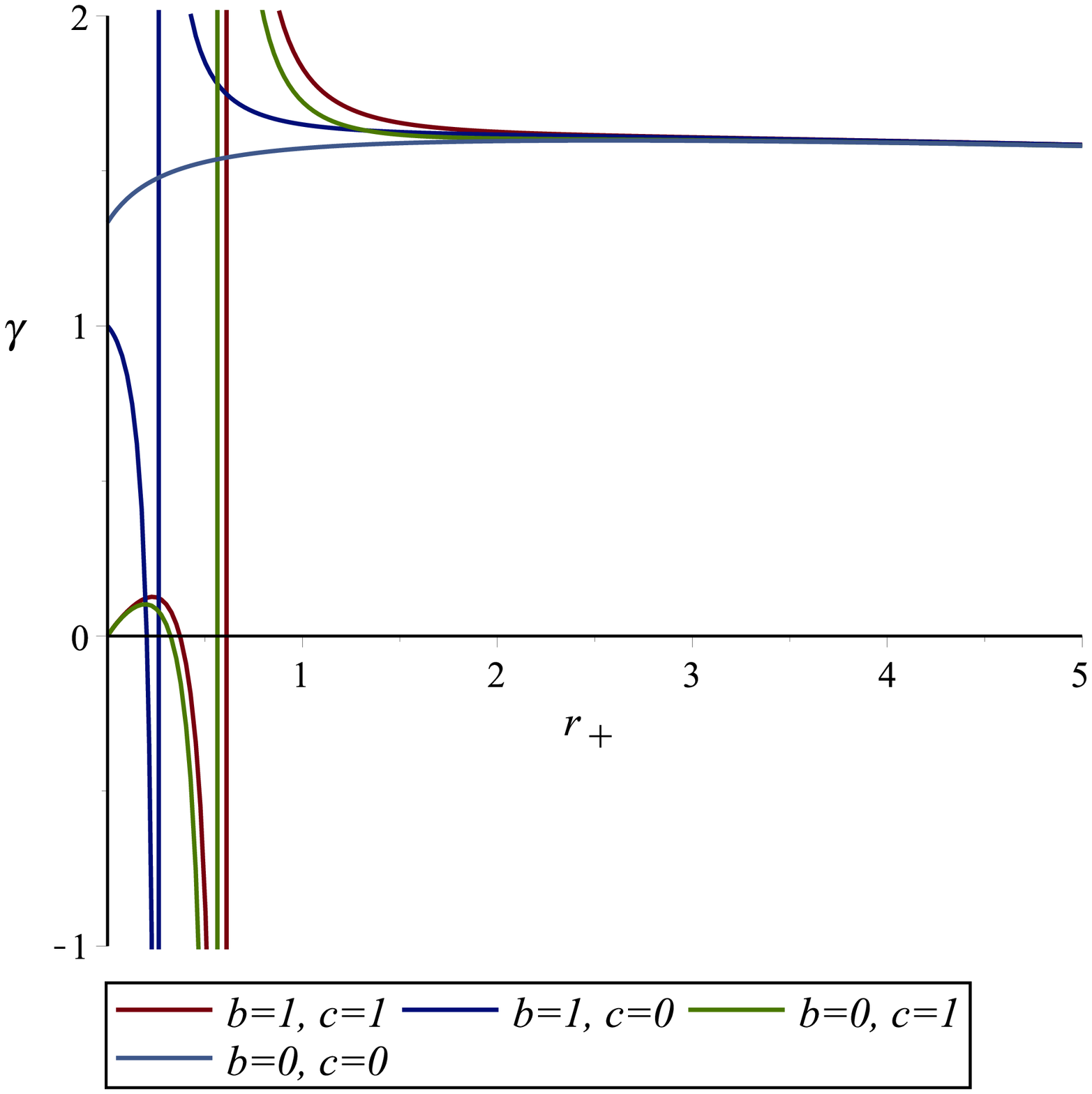}
    \end{tabular}
\caption{The plot of $\gamma$ for $f(R)$ BH for $\beta = 1.5$ and
$\Lambda = -0.1$ (left panel), $\Lambda = 0.1$ (right panel) }
  \end{figure}
\begin{eqnarray}\nonumber
  C_P|_{r=r_+} &=& -\frac{2}{3\pi r_+ (r_+^2 \Lambda +1)}\Big(-6\pi^2r_+^5\Lambda
  +10\pi^2\beta r_+^4 \\
  &+&4\pi b r_+^3 \Lambda-3\pi b \beta r_+^2+4\pi^2r_+^3
  +2cr_+\Lambda-2c\beta\Big).\label{17fa}
\end{eqnarray}
For this BH, the plot of $\gamma = C_p/C_v$ is shown in Fig.
\textbf{7}. We observe that the value of $\gamma$ increases for
larger horizon due to correction terms. Also, we find that the value
of $\gamma$ is same for both cases (positive and negative values of
$\Lambda$), i.e., $\gamma \rightarrow 0.6$ for larger horizon. Thus,
we can say that $\gamma$ shows similar stable behavior for both
cases of cosmological constant. It is realized that the value of
$\gamma$ becomes higher due to correction terms and exhibits similar
stable behavior for both cases of $\Lambda$ in $f(R)$ BHs.

\subsection{Phase transition}

\begin{table}[h]
\begin{center}
\begin{tabular}{|c|c|c|c|}
\hline
$\Lambda$&\begin{tabular}{@{}c@{}c@{}}correction\\
terms\end{tabular}&\begin{tabular}{@{}c@{}c@{}} range of \\local
stability\end{tabular}&phase transition\\
 \hline $-0.1$ &
\begin{tabular}{@{}c@{}c@{}c@{}}$b=1$, $c=1$ \\ $b=0$,
$c=1$\\ $b=1$, $c=0$\\$b=0$,
$c=0$\end{tabular}&\begin{tabular}{@{}c@{}c@{}c@{}}
$0<r_+<0.61,~r_+>3.2$\\ $0<r_+<0.27,~r_+>3.2$\\
$0<r_+<0.56,~r_+>3.2$\\$r_+>3.2$\end{tabular}&
\begin{tabular}{@{}c@{}c@{}c@{}c@{}}$0.61$\\$0.27$\\$0.56$\\$\phi$\end{tabular}\\
\hline 0.1&\begin{tabular}{@{}c@{}c@{}c@{}c@{}}$b=1$, $c=1$
\\
$b=0$, $c=1$
\\$b=1$, $c=0$\\$b=0$, $c=0$\end{tabular}&\begin{tabular}{@{}c@{}c@{}c@{}c@{}}
$0<r_+<0.61, ~r_+>30.33$ \\ $0<r_+<0.26,~r_+>30.33$\\
$0<r_+<0.56,~r_+>30.33$\\$~r_+>30.33$\end{tabular}&\begin{tabular}
{@{}c@{}c@{}c@{}c@{}}$0.61,~30.33$\\$0.26,~30.33$\\$0.56,~30.33$\\$30.33$\end{tabular}\\
\hline
\end{tabular}
\end{center}
\caption{Range of local stability and critical points of horizon
radius of phase transition for f(R) BH due to the effect of higher
order correction entropy}. \label{t2}
\end{table}
\begin{figure}
\centering
  \begin{tabular}{@{}cccc@{}}
    \includegraphics[width=.5\textwidth]{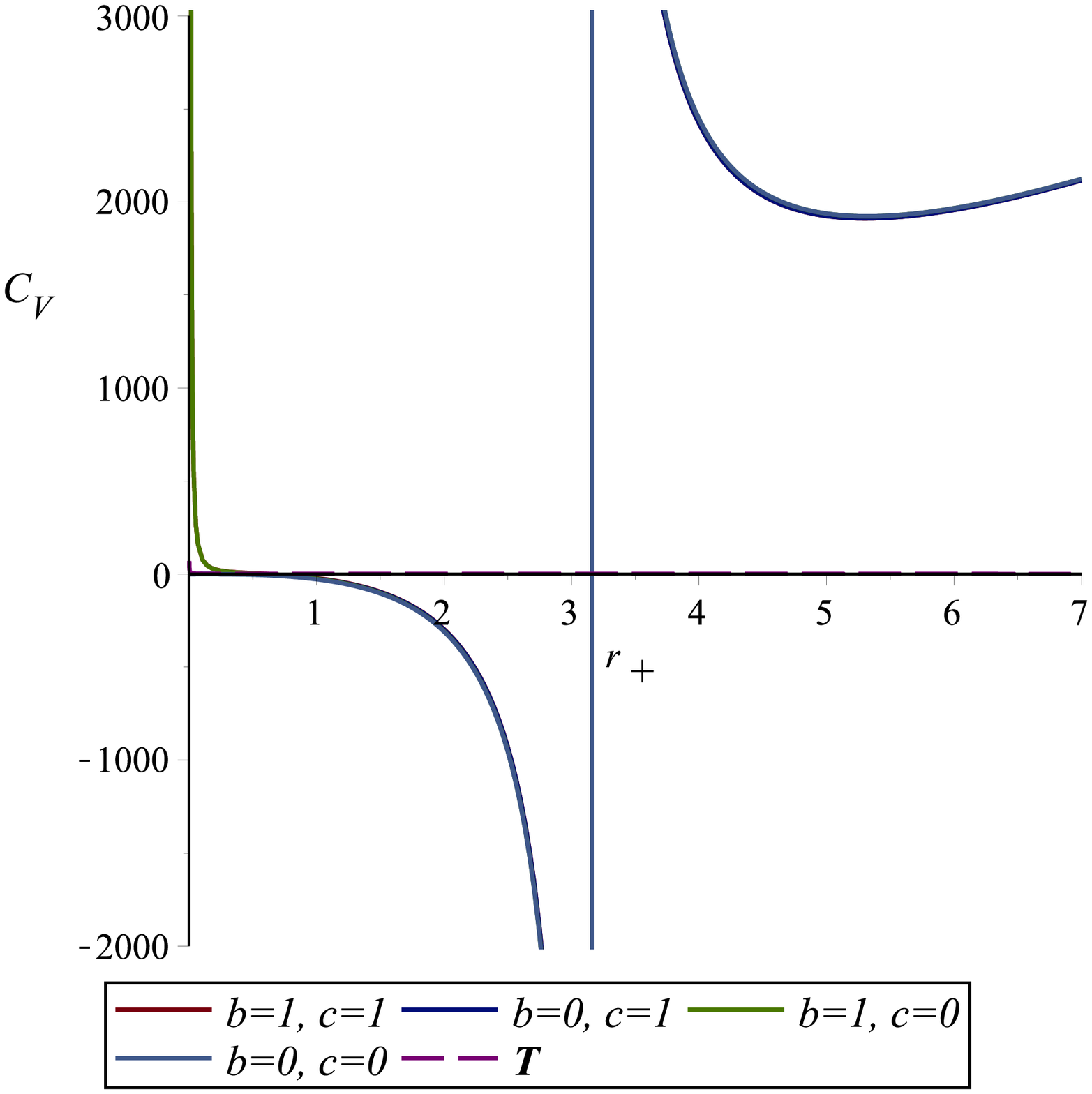} &
    \includegraphics[width=.5\textwidth]{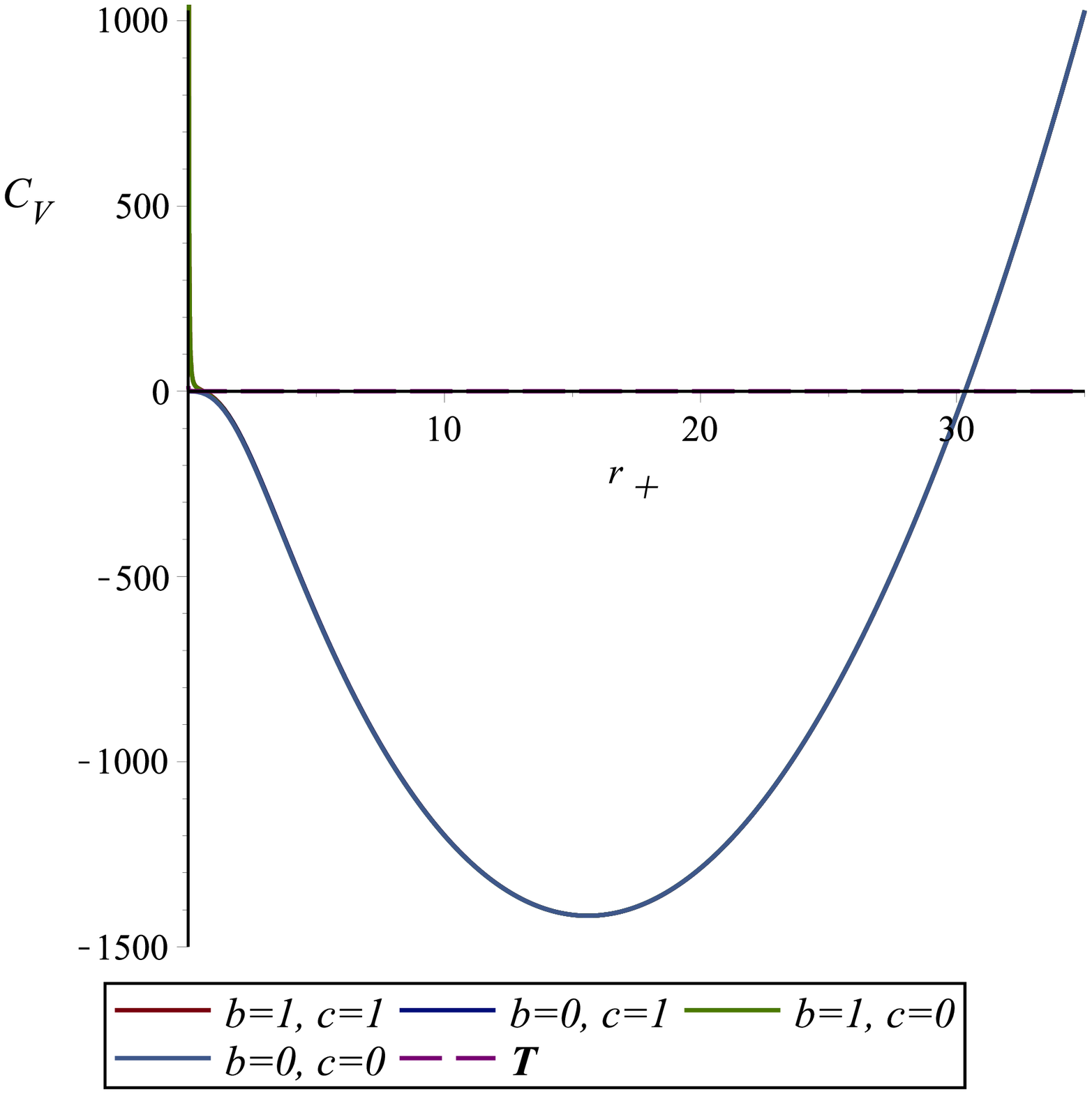}
    \end{tabular}
\caption{The plot of specific heat at constant volume for $f(R)$ BH
for $\beta = 1.5$ and $\Lambda = -0.1$ (left panel), $\Lambda = 0.1$
(right panel) }
  \end{figure}
We discuss the phase transition and range of local stability of BH
horizon due to thermal fluctuation for negative and positive values
of $\Lambda$. In both panels of Fig. \textbf{8}, it is observed that
the range of local stability is maximum for both correction terms
and then the range of local stability is higher for second order
correction term as compare to first order logarithmic correction
term. Furthermore, we also find the phase transition points for both
cases in Table \textbf{\ref{t2}}. We obtain more phase transition
points in the presence of correction term as compare to its absence.
Moreover, we have more phase transition points for positive
cosmological constant as compare to negative. We also observe that
$f(R)$ BH is the most locally stable for $r_+ > 3.2$ in the case of
negative $\Lambda$ while it is completely stable for $r_+ > 30.33$
in the case of positive $\Lambda$. Hence, we can conclude that if we
utilize the correction terms and increases the value of cosmological
constant then the range of local stability increases and we can
obtain more phase transition points.

\subsection{Grand Canonical Ensemble}

The free energy in grand canonical ensemble (Gibb's free energy) is
given by
\begin{eqnarray}\nonumber
  G|_{r=r_+} &=& -\frac{1}{24\pi^2r_+^3}\Big(2r_+^6\pi^2 \Lambda+ 6r_+^4 \pi^2+
  6cr_+^2\Lambda-12cr_+ \beta-6c\\
  &-&3br_+^2 \pi (r_+^2 \Lambda-2r_+ \beta-1)\ln \big(\frac{(-r_+^2\Lambda
  +2\beta r_+-1)^2}{16 \pi
}\big)\Big)-\mu r_+.\label{g1f}
\end{eqnarray}
\begin{figure}
\centering
  \begin{tabular}{@{}cccc@{}}
    \includegraphics[width=.5\textwidth]{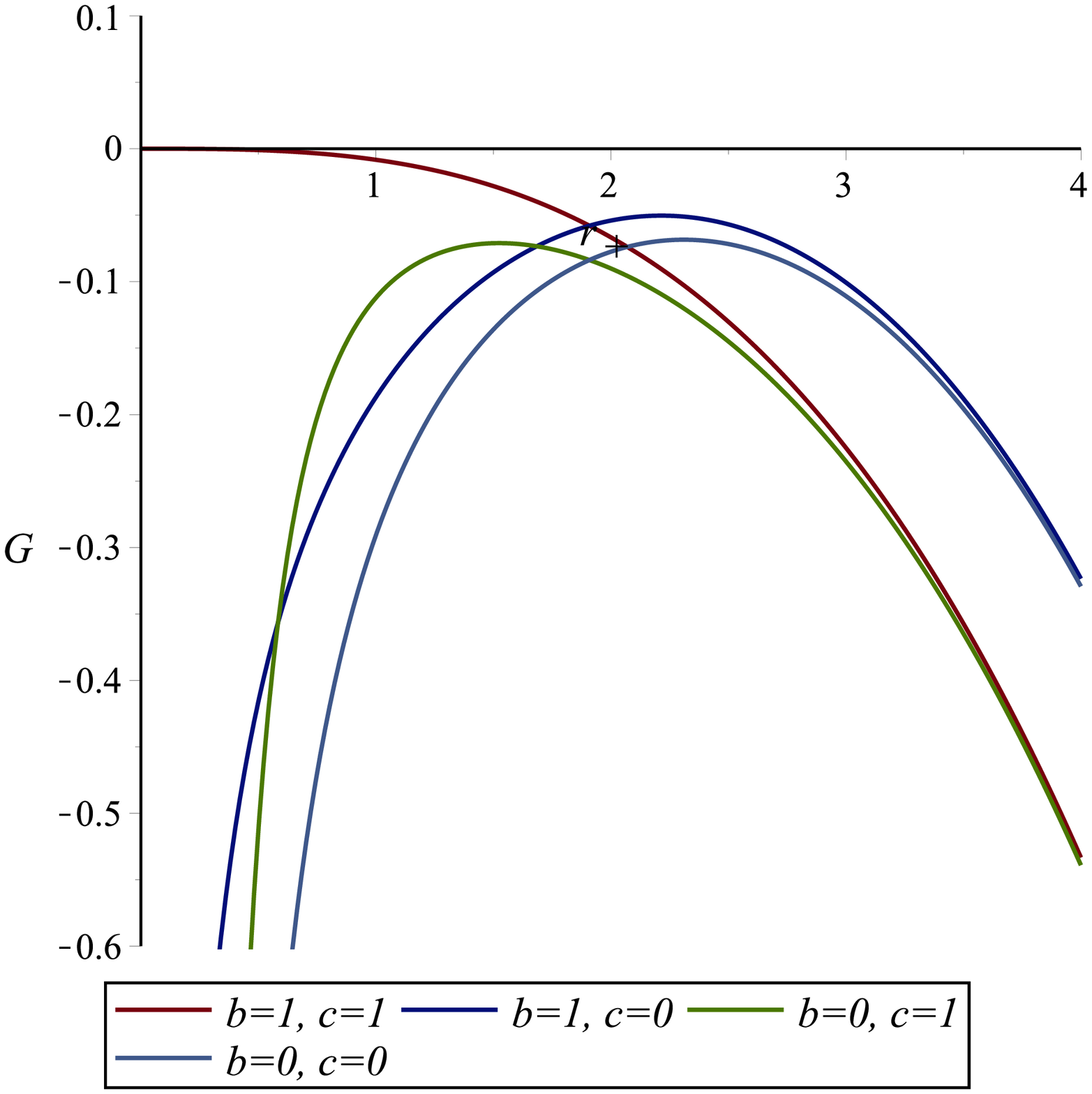} &
    \includegraphics[width=.5\textwidth]{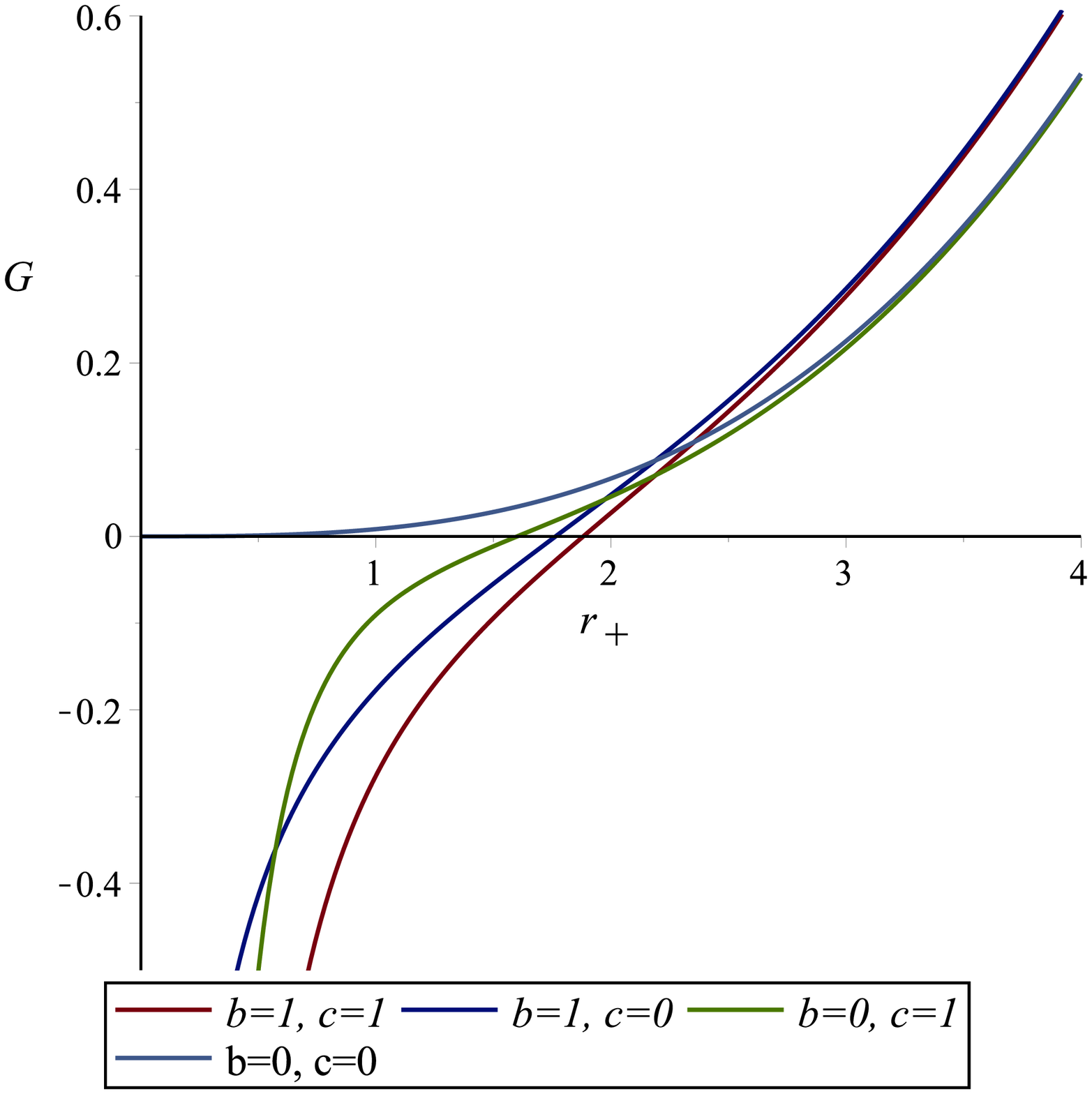}
    \end{tabular}
\caption{The plot of Gibb's free energy for $f(R)$ BH for $\beta =
1.5$ and $\Lambda = -0.1$ (left panel), $\Lambda = 0.1$ (right
panel) }
  \end{figure}
Fig. \textbf{9} represents the behavior of Gibb's free energy for
negative and positive cosmological constant. In both panels, it is
observed that the correction terms reduce the Gibb's free energy. In
left panel, we notice that the f(R) BH is stable for both correction
terms ($b=1, c=1$) as compare to others while at $r_+=1$, we find
the most stable BH by utilizing the logarithmic correction term
($b=1, c=0$). From right panel, we can see that Gibb's free energy
is higher for both correction terms and logarithmic correction term
as compare to others. Hence, we can conclude that $f(R)$ BH is the
most thermodynamically stable for higher order correction terms and
positive cosmological constant.

\subsection{Canonical Ensemble}
\begin{figure}
\centering
  \begin{tabular}{@{}cccc@{}}
    \includegraphics[width=.5\textwidth]{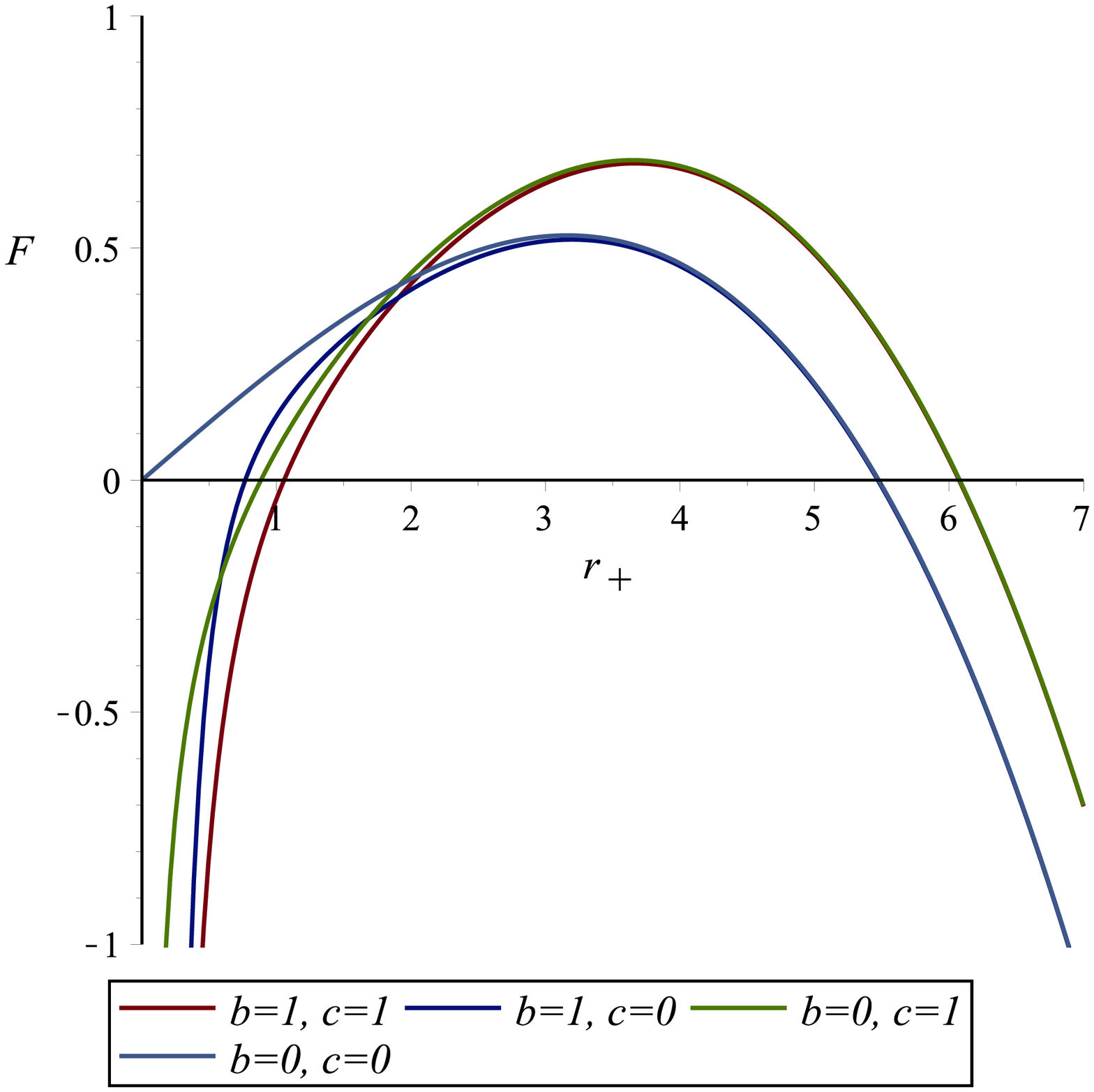} &
    \includegraphics[width=.5\textwidth]{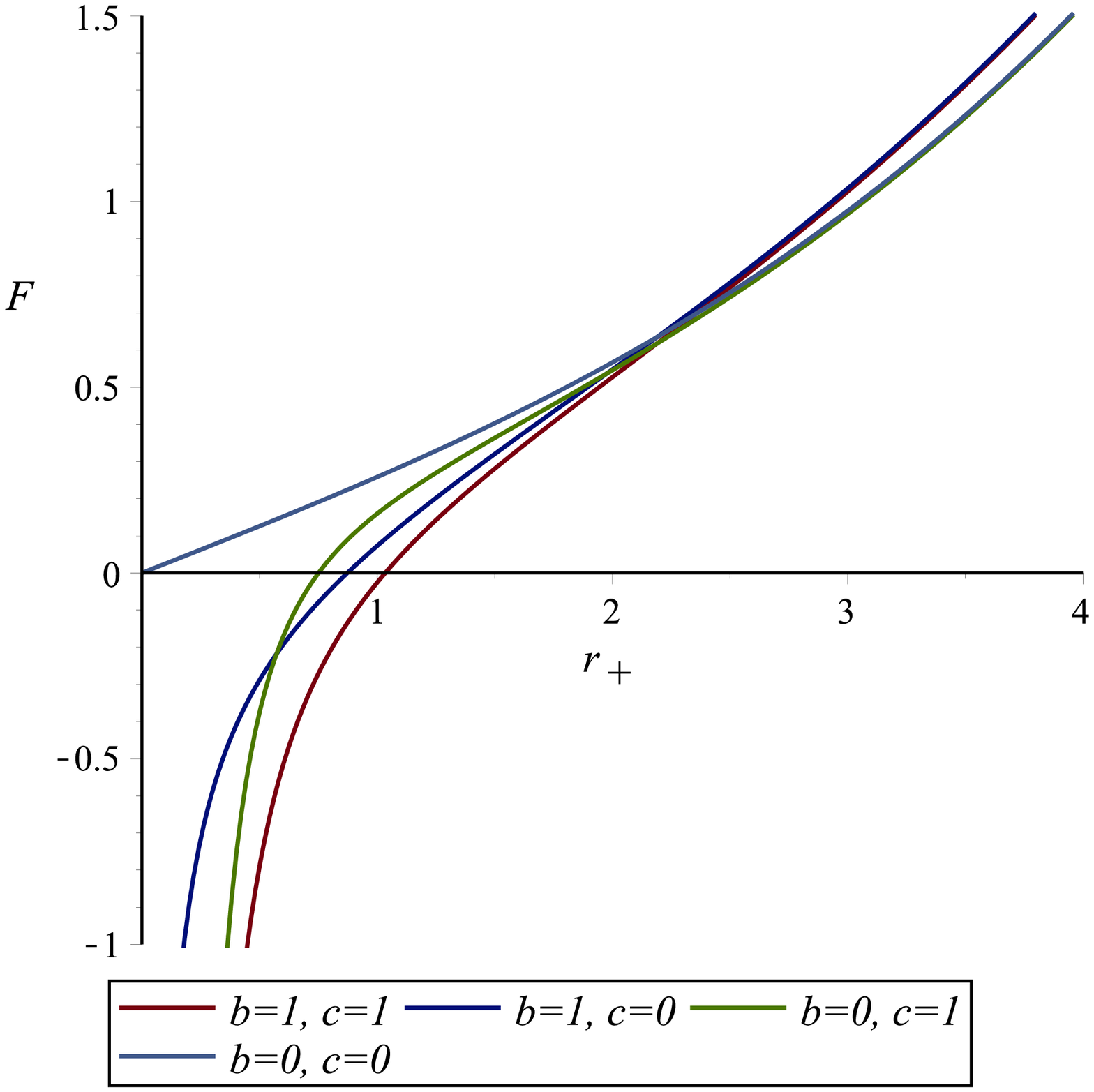}
    \end{tabular}
\caption{The plot of Helmhotz free energy for $f(R)$ BH for $\beta =
1.5$ and $\Lambda = -0.1$ (left panel), $\Lambda = 0.1$ (right
panel) }
  \end{figure}
 The free energy in canonical ensemble is
known as Helmhotz free energy if the charge is fixed, which is
\begin{eqnarray}\nonumber
  F|_{r=r_+} &=& -\frac{1}{24\pi^2r_+^3}\Big(2r_+^6\pi^2 \Lambda+ 6r_+^4 \pi^2+
  6cr_+^2\Lambda-12cr_+ \beta-6c\\
  &-&3br_+^2 \pi (r_+^2 \Lambda-2r_+ \beta-1)\ln \big(\frac{(-r_+^2\Lambda
  +2\beta r_+-1)^2}{16 \pi
}\big)\Big)\label{f1f}
\end{eqnarray}
The behavior of Helmhotz free energy for negative and positive
cosmological constant is plotted in Fig. \textbf{10}. In both
panels, we observe that free energy decreases due to correction
terms. In left panel, one can see the free energy is positive till
$r_+=6$ and the BH is most stable for both correction term (red
line) and second order correction term (green line) in the case of
negative cosmological constant. In right panel, for positive value
of $\Lambda$, BH is most stable for both correction term (red line)
and logarithmic correction term (blue line) as compare to others.
Hence, we can conclude that $f(R)$ BH shows most stable behavior for
positive values of cosmological constant and higher order correction
terms.

\section{Conclusion}

In this paper, we considered the RN-AdS BH with GM and $f(R)$ BH,
and discussed the thermodynamics in the presence of higher order
corrections of entropy. We utilize the results of corrected entropy
by setting first order term is logarithmic and second order term is
inversely proportional to original entropy. In these scenarios, we
have studied the behavior of pressure and specific heat for both
BHs. It is observed that the pressure reduces for second order
correction term for both BHs but the pressure of RN-AdS BH with GM
also decreases by considering higher values of cosmological constant
while there is no change in pressure with respect to cosmological
constant in $f(R)$ BH. We have also discussed the ratio of specific
heat ($\gamma$) at constant pressure and volume for both BHs. We
have observed that the value of $\gamma$ increases due to the effect
of higher order correction terms for both BHs. The value of $\gamma$
for RN-AdS BH with GM exhibits more stable behavior for lower values
of cosmological constant while $f(R)$ BH shows similar behavior for
both (negative and positive) cases of $\Lambda$.

We have also studied the phase transition due to the effect of
higher order correction terms for both BHs and obtained the phase
transition points. We have noticed that the range of local stability
and the phase transition points increases due to the effect of
second order correction term and the higher values of cosmological
constant for both BHs. We have investigated the free energy in
canonical (Helmhotz free energy) and and grand canonical (Gibb's
free energy) ensembles. We have observed that the free energy
reduces due to higher order correction terms. The Helmhotz free
energy and Gibb's free energy show the most stable behavior for both
BHs in case of positive cosmological constant as compare to negative
cosmological constant. We noticed that the both free energies
exhibit the most stable behavior by utilizing the both correction
terms ($b=1, c=1$). The Gibb's free energy is reduced in both BHs
due to chemical potential as compare to Helmhotz free energy. We
also observe that both BHs exhibit the most locally stable behavior
for negative cosmological constant while both BHs show the most
globally stable behavior for positive values of cosmological
constant. Hence, it is concluded that both BHs show the most locally
stable behavior for second order correction term ($b=0, c=1$) while
both BHs is most globally stable by utilizing the both correction
terms ($b=1, c=1$), so it is better to consider the higher order
correction terms.

\end{document}